\documentclass[draft]{agujournal2019}
\usepackage{url} 
\usepackage{lineno}
\usepackage[inline]{trackchanges} 
\usepackage{soul}
\draftfalse

\journalname{Earth and Space Science}

\begin{document}

%
%


\title{Revisiting the limits of atmospheric temperature retrieval from cosmic-ray measurements}

%
%




\authors{Irma Ri\'adigos\affil{1,2}, Diego Gonz\'alez-D\'iaz\affil{2}, and Vicente P\'erez-Mu\~nuzuri\affil{1}}


\affiliation{1}{CRETUS Institute, Group of NonLinear Physics, Faculty of Physics, University of Santiago de Compostela, Spain}
\affiliation{2}{Instituto Galego de F\'isica de Altas Enerx\'ias (IGFAE), Faculty of Physics, University of Santiago de Compostela, Spain}




\correspondingauthor{Irma Ri\'adigos}{irma.riadigos@usc.es}




\begin{keypoints}
\item Improved approach to obtain atmospheric temperature from cosmic-ray data by introducing a pair of detection stations and angular information 
\item Results show a significant retrieval improvement over earlier methods with high accuracy up to 20 km of height
\item The use of affordable and small-scale cosmic-ray telescopes is suggested as a possible complement to satellite temperature measurements
\end{keypoints}

%
%

%
%


\begin{abstract}
A priori, cosmic-ray measurements offer a unique capability to determine the vertical profile of atmospheric temperatures directly from ground. However, despite the increased understanding of the impact of the atmosphere on cosmic-ray rates, attempts to explore the technological potential of the latter for atmospheric physics remain very limited.
In this paper we examine the intrinsic limits of the process of cosmic-ray data inversion for atmospheric temperature retrieval, by combining a detection station at ground with another one placed at an optimal depth, and making full use of the angular information. With that aim, the temperature-induced variations in c. r. rates have been simulated resorting to the theoretical temperature coefficients $W_T(h, \theta, E_{th})$ and the temperature profiles obtained from the ERA5 atmospheric reanalysis.
Muon absorption and Poisson statistics have been included to increase realism. The resulting c.r. sample has been used as input for the inverse problem and the obtained temperatures compared to the input temperature data. Relative to early simulation works, performed without using angular information and relying on underground temperature coefficients from a sub-optimal depth, our analysis shows a strong improvement in temperature predictability for all atmospheric layers up to 50 hPa, nearing a factor 2 error reduction. Furthermore, the temperature predictability on 6~h-intervals stays well within the range 0.8-2.2~K. Most remarkably, we show that it can be achieved with small-area m$^2$-scale muon hodoscopes, amenable nowadays to a large variety of technologies. For mid-latitude locations, the optimal depth of the underground station is around 20~m.
\end{abstract}

\section*{Plain Language Summary}
Cosmic rays (c.r.) are a form of natural radiation that comes from outer space and traverses the atmosphere. Analogously to X-ray radiation, we can extract information from the object they pass through, the atmosphere in this case. Cosmic radiation can be measured at the Earth's surface using sophisticated instruments. In our work we examine in detail the possibility of retrieving the vertical profile of temperatures from c.r. measurements, by using two detection stations, one placed at the surface and another one at an optimal depth. We make full use of a feature which is characteristic of modern c.r. detectors, the angular information. This refers to the capability of measuring c.r. from different directions. To analyze the limits of the suggested approach, we estimate temperatures from simulated c.r. data that would be measured under realistic atmospheric conditions and compare them with the original ones. Relative to early simulation works, our method estimates temperatures with a greater accuracy at higher temporal resolutions and for atmospheric layers up to ~20 km. Most remarkably, we show that this performance can be achieved with small and affordable detectors, bringing the possibility of complementing satellite-borne temperature retrieval with a technology cheaper to assemble and maintain.

%
%

%


%
%
%
%

\section{Introduction}

Since its discovery in 1912, cosmic-ray radiation has offered an exceptional way to observe the world around us from different points of view. Cosmic rays (c.r.) consist of high energy particles whose origin lies in outer space, traveling at nearly the speed of light. Chiefly composed of charged particles (protons), they are continuously reaching the Earth from all directions. As soon as this radiation arrives at our planet, it stumbles upon two shields: the Earth's magnetic field and the atmosphere. The first one deflects the less energetic particles preventing them from entering the atmosphere. However, if they are able to pass this barrier they will eventually collide with the atmospheric nuclei, thus starting a cascade of secondary particles. A chain reaction gets underway, generating more secondary particles until they reach the ground. This cascade (often dubbed ``air shower'') encompasses common subatomic particles such as protons, neutrons, pions, kaons, muons, electrons, positrons, gamma-rays and neutrinos. Muons ($\mu^{+/-}$), highly penetrating lepton-particles similar to electrons but heavier, are among the most numerous secondary particles at sea level and they can be measured using detectors on the surface and even at great underground depths of several hundred meters.
This radiation traverses long distances in the atmosphere, analogously to what X-rays do in low-$Z$ materials, providing integrated information about its internal structure (equivalently, its mass density as a function of depth). The essential aspects of the process can be found, for instance, in \citeA{dorman2004}.

Major factors influencing c.r. rates at ground are surface pressure and the temperature profile of the atmosphere \cite{sagisaka1986}. The former, responsible for the so-called ``barometric effect'', varies in anti-correlation with c.r. rates. In broad terms, the effect stems from the increased opacity of the atmosphere: a higher surface pressure implies a greater amount of air mass traversed, with less particles reaching ground.
The ``temperature effect'', on the other hand, results from local variations of the air density caused by temperature changes, whose final impact is accumulated over the particle's path through the atmosphere. 
When a specific layer of the atmosphere becomes colder, for instance, its density increases. Therefore, particles will more often interact rather than decay since their interaction probability is proportional to the density of scatter centers. By way of illustration, muons originate mainly when charged pions and kaons ($\pi^{+/-}, \textnormal{K}^{+/-}$) decay and so, as the temperature increases and their interaction probability consequently decreases, decaying to the more penetrating muons becomes a more likely pathway. Considering this effect alone, c.r. rates at ground would increase with increasing temperature \cite{duperier1949positive}. Even though daughter muons (due its leptonic nature) are far less interacting than their pion/kaon parents (hadronic particles), they are unstable and can decay to the lightest leptons in their family: electrons/positrons. Therefore, for a complete picture of the temperature effect it must be taken into account that, as the atmosphere expands due to a temperature increase, muons will have to travel greater distances to reach ground, increasing their probability of decaying in-flight. The (lighter) electrons and positrons will be much more readily absorbed than muons themselves through ``bremsstrahlung'' losses, a physical process that is strongly enhanced for low-mass particles. As a result of this compensatory effect, c.r. rates would decrease as the temperature increases \cite{sagisaka1986}. This latter effect is very strong in particular for low-energy muons (that dominate the flux at ground level), given that the muon lifetime becomes shorter with energy as a consequence of relativistic time dilation.


\begin{figure}[h!!]
    \centering
    \includegraphics[scale=0.55]{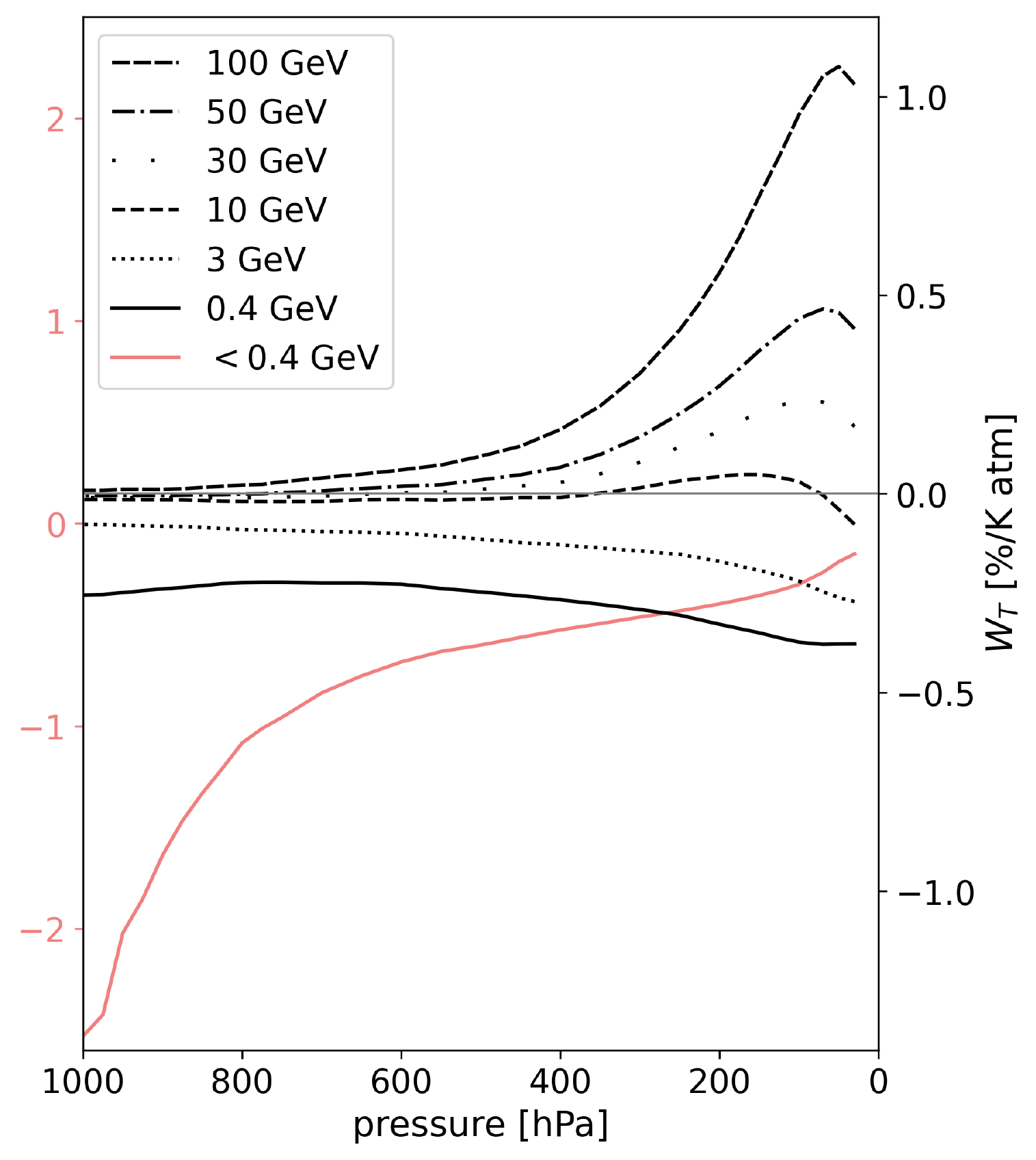}
    \caption{Temperature coefficients for vertical direction ($\theta$=0$^\circ$) at several threshold energies corresponding to different underground depths \cite{sagisaka1986}, shown in black on the right axis. Coefficients for vertical muons observed at ground, tagged respectively by passage or absorption in 10~cm-lead and labelled as ``hard'' ($E_{th}= 0.4$ GeV) and ``soft'' ($E< 0.4$ GeV) are shown by continuous lines (black and red, respectively). The latter have been obtained from \cite{dorman2004}, and have axis on the left.}
    \label{fig:temperature_coefficients}
\end{figure}

\begin{figure}[h!!]
\centering
\includegraphics[width=0.75\textwidth]{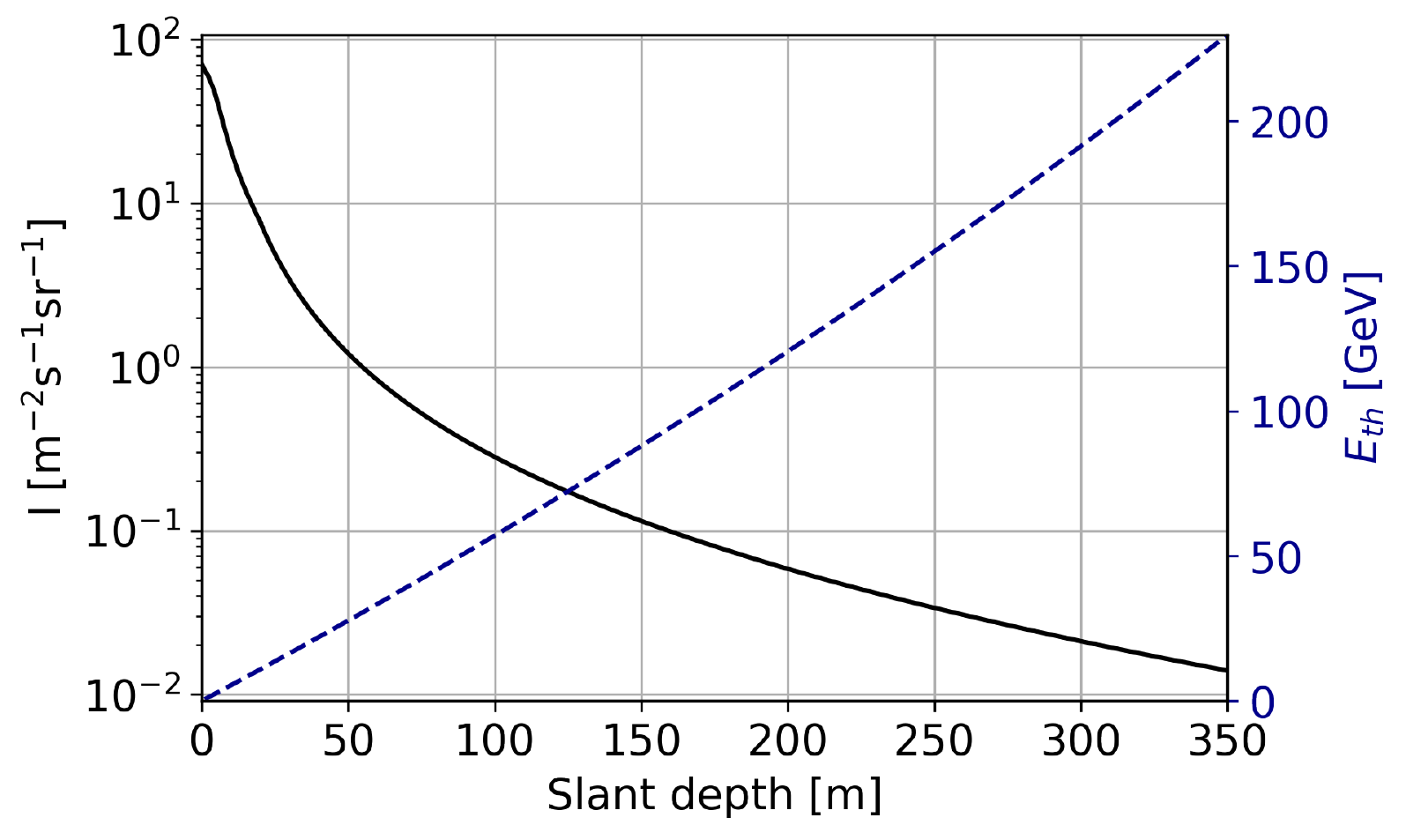}
\caption{Parameterizations used in this work: vertical muon intensity (left axis) and threshold energy (right axis) as a function of vertical slant depth. A typical soil density of $\rho = 2.7$ g/cm$^3$ has been assumed.}
\label{fig:intensity_vs_depth}
\end{figure}

The well-established aforementioned phenomena are known as the ``negative effect'' (dominant at ground level) and the ``positive effect'' (dominant in underground stations). Since the temperature profile of the atmosphere has a complex behavior, and the impact on c.r. rates is accumulative along the particles trajectories, it is customary to describe these effects through the so-called temperature coefficients, $W_T(h, \theta, E_{th})$. They are given as a function of the atmospheric height ($h$), depending on the angle of incidence ($\theta$) and the threshold energy ($E_{th}$) of the muons (equivalently, the underground depth at which observations are made, see Figure \ref{fig:intensity_vs_depth}). Theoretical calculations for $W_T$ became widespread with L. I. Dorman (for a recent compilation see \citeA{dorman2004}, \citeA{sagisaka1986} and, more recently, \citeA{dmitrieva2011coefficients}). With the aim to illustrate the main effects, they have been reproduced in Figure \ref{fig:temperature_coefficients} for the case of vertical incidence. The coefficients are normalized in terms of standard ``atm'' pressure (1 atm = 1013 hPa), as given by Dorman and Dmitrieva \textit{et al.} We provide the \textit{x}-axis in units of hPa for better comparison with the results in the following sections.

The temperature coefficients (given in units of \%/K atm) indicate how much rates at observation level $X$ vary for each degree of temperature change at the specific atmospheric depth $h$. The coefficients are the sum of two terms: the positive effect related to mesons and the negative effect associated with muons.The sign of the coefficient gives information about the net value of the total effect for each for each layer, which depends on $E_{th}$. In the case of low threshold energies, the negative effect dominates, which means that for every temperature increase in that layer, the measured change in rates will have the opposite sign. However, as the threshold energy increases, the positive effect prevails (see Figure \ref{fig:temperature_coefficients}). The latter effect becomes prominent around 100 hPa ($\sim$15 km) where the peak of meson production takes place. This can be appreciated in Figure 1 for high threshold energies. On the other hand, the negative effect associated to muons has to do with the decrease in surviving muons produced at a certain depth, as mentioned above.

Therefore, an analogy can be drawn between satellite and cosmic-ray measurements when targeting atmospheric temperature retrieval. Weather satellites employ the observations of electromagnetic radiations emitted by the atmosphere, that depend on its state. Temperature-dependent weighting functions, based on competing emission and absorption processes, need to be established before-hand for each atmospheric layer. Besides, they depend on the energy of the measured radiation \cite{eyre1991satellites}. The inverse problem of retrieving the atmospheric temperature profile can be solved by combining several energy channels. By the same token, c.r. measurements at different angles and energies (ground/underground) may be used for the same purpose. 
As seen in Figure \ref{fig:temperature_coefficients}, selection by threshold energy (equivalently, depth) is the most natural way to separate the positive and negative effects in the weights, allowing a priori a higher sensitivity to the behaviour of the different atmospheric layers.

One of the first attempts to estimate the atmospheric temperature profile using c.r. measurements with multiple detectors was carried out by \citeA{kohno1981estimation}. By combining measurements at ground for soft muons (stopped in 10~cm of plastic scintillator), hard muons (passing through 10~cm of lead) and underground muons (80~mwe) performed with 1m$^2$-area detectors they claimed a daily accuracy in the range 2-2.5~K for the atmospheric regions corresponding to 100, 500 and 900~hPa, over a period of about half a year. The detector suite was accompanied by measurements from a neutron monitor, in order to independently identify solar or interplanetary events that could bias the temperature estimate. Despite the sophistication of the approach, neither the depth of the detector was optimized nor angular information was used. In the light of those very promising results, it is surprising to these authors that such natural extensions were not pursued. In fact, more recent studies have endeavored to obtain the temperature for more atmospheric layers by using a single detector and measurements at different angles (for example, \citeA{yanchukovsky2007,yanchukovsky2015,yanchukovsky2020}). 

Clearly one early limitation was detector complexity, as measurements were done over large areas by resorting to large scintillator tiles. However, multi-directional muon detectors (sometimes called hodoscopes) are nowadays common-place and available as part of the Global Muon Detector Network (GMDN) \cite{GMDNrockenbach}, for instance. Furthermore, the revival of the fields of muon tomography \cite{procureur2018muonimaging} and muography \cite{pyramids2017morishima} has led to the adaptation of new technologies from particle physics, including for instance the development of extruded plastic scintillator \cite{pla2003extruding}, micropattern gaseous detectors \cite{gomez2019muon,particles4030028}, as well as classic \cite{baesso2014toward} and timing \cite{xing2014position}  resistive plate chambers (RPCs), just to name a few. They all offer affordable ways to cover large areas at high angular resolution. As an example, the effective atmospheric temperature has been recently measured at ground with a 2~m$^2$ timing RPC station in \cite{riadigos2020atmospheric}, the first time that this technology, capable of time resolutions down to 50-60~ps and precise angular reconstruction on areas of several m$^2$ \cite{watanabe2019compensated,blanco2020ship}, has been used for the task. In view of these powerful technological assets, the existence of new detailed calculations of the atmospheric weights, as well as the latest generation of accurate temperature data from the European Centre for Medium-Range Weather Forecast (ECMWF), reassessing the technological potential of cosmic rays for atmospheric temperature forecast seems very timely if not imperative.

\section{Temperature effect}

Even for a perfect detector, cosmic-ray rates are subject to variations of diverse origins: those due to changes in the solar activity and in the atmosphere thermodynamic state are the most important. The former act as a potential systematic bias to the atmospheric temperature estimate, and in the remainder of this work we will assume implicitly that they can be isolated and eliminated. Although a natural way to perform this task is through the complementary use of neutron detectors (highly insensitive to atmospheric temperature variations), underground muon detectors (as the one proposed in text) may be sufficient, as solar and interplanetary events have less influence at underground depths. This occurs because interplanetary phenomena affect low-energy primary cosmic rays (in the range of a few MeV and GeV), which in turn are responsible for originating the low-energy secondary muons at sea level \cite{gaisser1991,dorman2004}. Along these lines, changes in the measured c.r. rates induced by temperature variations can be approximated by the following expression \cite{dmitrieva2011coefficients}:
\begin{equation}
    \frac{\Delta R}{R_0} (E_{th}, \theta)= \sum_{i=1}^{n} W_T (E_{th},\theta, h_i) \Delta T_i \Delta h_i
\label{eq:temperature_effect}
\end{equation}
Here $\Delta R/R_0$ is the relative variation of c.r. rates at zenith angle $\theta$ and for a muon threshold energy $E_{th}$, with $R_0$ representing the average rate over a particular period of time; $W_T (E_{th},\theta, h_i)$ are the temperature coefficients, specific for a certain atmospheric layer $i$ at height (pressure) $h_i$;  $\Delta T_i=T_i-T_{i,0}$ represent the temperature variations of a layer with respect to its mean value $T_{i,0}$; $\Delta h_i=h_{i-1}-h_i$ refers to the layer thickness (in units of pressure). 

\section{Methods}

\subsection{Cosmic-ray intensity simulation}\label{CRsimul}

For a mid-latitude location around 40$^\circ$, the intensity of vertical muons at ground is typically $I_g \approx 70$ m$^{-2}$s$^{-1}$sr$^{-1}$ \cite{grieder2001cosmic,haino2004surfmuons}, the value used hereafter. In this case, the angular dependence has been parameterized as $dN/d\cos(\theta) \propto (\cos {\theta})^2$ for either soft ($<0.4$~GeV) or hard ($>0.4$~GeV) muons \cite <e.g.,>[]{reyna2006simple}. Given the very high statistics for any angular bin, the particular choice of this distribution does not influence the temperature retrieval for muons reconstructed at ground level.

In the case of underground measurements, \citeA{lipari1991} have shown a simple relation between muon intensity and slant depth, $X$:
\begin{equation}
    I(X) = I_u \left( \frac{X_0}{X}\right)^{\eta} e^{-\frac{X}{X_0}}
\label{eq:depth-intensity}
\end{equation}
where $I_u$ = 2.15 $\times$ 10$^{-6}$ cm$^{-2}$s$^{-1}$sr$^{-1}$, $\eta$=1.93 and $X_0$=1155 mwe \cite{aglietta1998depthintensity}. Since this parameterization is not accurate for shallow depths ($<20$~m) it has been complemented here by the one presented in \citeA{bogdanova2006cosmic}, which is obtained from an approximation of the surface muon spectrum together with muon range tables. Figure \ref{fig:intensity_vs_depth} shows the muon intensity (in units of m$^{-2}$s$^{-1}$sr$^{-1}$) as a function of slant depth using the aforementioned parameterizations for an average soil density of 2.7 g/cm$^3$.

In addition, the threshold energy as a function of $X$ is given by \cite{gaisser1991}:
\begin{equation}
  E_{th} = \epsilon (e^{X/b}-1)  
\label{eq:eth-depth}
\end{equation}
where $\epsilon \sim$ 500 GeV and $b \sim$ 2.5$\times$10$^5$ g/cm$^2$ for muons in rock. For the simulation of underground muons we have assumed in the following an isotropic distribution impinging on a homogeneous soil slab, with rates and threshold energies for each angle obtained from eqs. \ref{eq:depth-intensity} and \ref{eq:eth-depth}.

For a given detector with a fixed detection area ($A$), the number of counts  ($N$) measured over a period of time ($\Delta t$) and solid angle ($\Delta \Omega$) is:
\begin{equation}
    N = I \cdot A \cdot \Delta t \cdot \Delta \Omega ~~~~~ (\equiv R_0 \Delta t)
\label{eq:counting_rates}
\end{equation}
and, for simplicity, the detection efficiency and acceptance have been assumed to be angle-independent and close to 1. The statistical fluctuations associated to $N$ are then given by $\sqrt{N}$. On account of that, the variations of c.r. rates can be re-expressed as:
\begin{equation}
    \left(\frac{\Delta N}{N_0} \right)_{obs} \equiv \left(\frac{\Delta R}{R_0} \right)_{obs} = \left(\frac{\Delta R}{R_0} \right)_{T} + \left(\frac{\Delta R}{R_0} \right)_{err}
\label{eq:temp_stat_changes}
\end{equation}
where $(\Delta R/R_0)_{obs}$ are the experimental c.r. rate variations, $( \Delta R/R_0 )_{T}$ the changes due to the temperature effect (Eq. \ref{eq:temperature_effect}), and $( \Delta R/R_0 )_{err}$ the associated statistical fluctuations of the measurement. Specially underground, the detector area and measuring time become critical variables. Based on a preliminary analysis, and practical considerations, we set for a detector area of 4 m$^2$, and a time interval of 6~h, although the impact of these choices in our analysis is evaluated at the end of the work.

In sum, the procedure followed for the simulation of c.r. variations over a specific period of time can be sketched as:
\begin{itemize}
    \item For underground detectors, c.r. intensity and threshold energy are estimated using Eq. \ref{eq:depth-intensity} and \ref{eq:eth-depth}, respectively, for a certain slab thickness over the detector (depth). The approximation of \citeA{bogdanova2006cosmic} is used to calculate intensities for depths shallower than $20$~m. Isotropic emission at ground level is assumed, for muons reaching underground.
    \item For surface detectors, a vertical c.r. intensity of $I_g \sim 70$ m$^{-2}$s$^{-1}$sr$^{-1}$ is assumed, following an angular distribution like $dN/d\cos(\theta) \propto \cos(\theta)^2$. For the soft component, the value of the intensity is assumed to be about a fraction 0.4 of the hard component \cite{dorman2004}.
    \item Counting rates, $N$, are calculated for a fixed detector size, time interval and solid angle as indicated in Eq. \ref{eq:counting_rates}.
    \item Variations in c.r. rates due to the temperature effect are calculated with Eq. \ref{eq:temperature_effect} using the temperature time series from ERA5 and the temperature coefficients for the corresponding energy, $E_{th}$, and angle, $\theta$ (linearly interpolated when needed). A more detailed discussion of the estimation of these coefficients is presented below in Section \ref{sec:temp_coeff}.
    \item Poisson noise with a mean value of $N$ was added.
\end{itemize}

\subsection{Temperature data}

Vertical profiles of atmospheric temperatures were retrieved from ECMWF reanalysis, using the ERA5 dataset which offers 37 isobaric levels (1000, 975, 950, 925, 900, 875, 850, 825, 800, 775, 750, 700, 650, 600, 550, 500, 450, 400, 350, 300, 250, 225, 200, 175, 150, 125, 100, 70, 50, 30, 20, 10, 7, 5, 3, 2, and 1 hPa), with a horizontal spatial resolution of 0.25$^\circ$ and a temporal resolution of 6 h \cite{ERA5}. A mid-latitude location at 40$^\circ$ was chosen, in this case corresponding to Santiago de Compostela (Spain). 

\subsection{Temperature coefficients}\label{sec:temp_coeff}
The distributions of temperature coefficients ($W_T$) have been calculated before by several authors. Dorman supplied the most extensive calculations for a wide variety of threshold energies and zenith angles \cite{dorman1972}. These estimates were later re-evaluated by Sagisaka and Dmitrieva \cite{sagisaka1986,dmitrieva2011coefficients}. The former provided coefficients for various combinations of threshold energy and zenith angle whereas the latter introduced up-to-date parameters in the calculations to give a vast database of coefficients, focused on threshold energies for surface hodoscopes. In the following, Dorman's coefficients will be used whenever the soft component at ground is involved, as he is the only one to provide them; underground coefficients for different angles will be taken from Sagisaka's work; for the hard component at ground, the coefficients supplied by Dmitrieva et al. will be adopted. In this latter case, a back-to-back comparison with Dorman and Sagisaka's weights is possible and will be discussed in text. Illustratively, a compilation of the atmospheric weights in case of vertical incidence is shown in Figure \ref{fig:temperature_coefficients}. They will be used for the single-channel (single-angle) analysis in section \ref{s-c}. For stations with angular resolution, weights as a function of angle are needed, and they will be introduced in section \ref{m-c}.

\subsection{Formulation of the inverse problem for temperature retrieval from cosmic-ray data}

The inverse problem of estimating the vertical distribution of atmospheric temperature from c.r. observations will be performed in this work through a simple linear regression. Thus, the temperature at the $i$-th layer of the atmosphere can be estimated from the c.r. rate variations as:
\begin{equation}
\Delta \hat{T}_i   = \sum_{k=1}^{n_{st}} \sum_{j=1}^{n_{ch}} c_{jki} \frac{\Delta R}{R_0}(E_{th,jk}, \theta_j) \bigg|_{k}
\label{eq:inverse_problem}
\end{equation}
where $c_{jki}$ are the coefficients of the least squares minimization and $\frac{\Delta R}{R_0}(E_{th,jk}, \theta_j) \big|_{k}$ are the relative variations of c.r. rates for a certain detector station $k$ (if several assumed) and angular bin $j$, with the threshold energy taken to be different for each station and angular bin (for stations detecting soft muons the threshold energy must be understood as a maximum energy). Rates have been simulated according to the procedure described in subsection \ref{CRsimul}. 

It must be noted before the start that, to date, various underground experiments have attempted to retrieve the temperature of the stratosphere from c.r. data, a particularly attractive possibility as the underground component is strongly influenced by the temperature of the upper atmospheric layers \cite{barrett1952effecttemp,ambrosio1997macro,osprey2009suddenminos,tilav2010}. Nevertheless, the statistical fluctuations in the counting process as well as the need of deep underground facilities play an important role that limits the practical use of this approach for atmospheric physics. Rate variations due to the temperature effect when deep underground are indeed around $\sim$1.5 \%, and therefore statistical fluctuations need to be much lower than this in order to retrieve information about the atmosphere. By way of illustration, the MINOS experiment (located 720 m underground) has a mean counting of $\sim$40000 muons per day that results in error bars of the order of just 0.5 \%, thanks to an imposing acceptance of 691~m$^2$sr. In this way, seasonal rate variations caused by the temperature effect could be observed in \citeA{adamson2010minos}. On the other end, a dedicated campaign carried out at the Canfranc Underground Laboratory characterized the muon flux inside the experimental halls with a muon monitor of just 0.95 m$^2$ \cite{trzaska2019canfranc}. Despite the great precision achieved in the reconstruction of the flux as a function of azimuth and zenith angles, with $\sim$400 muons per day and a statistical uncertainty of $\sim$5 \%, observing the temperature effect became impossible. Figure S1 in Supplementary Information provides additional details of the limitations of this approach.  

The above facts highlight the main limitations of a single-channel (single-angle) analysis, therefore an alternative approach to the inversion of cosmic-ray data is proposed in this work, based on eq. \ref{eq:inverse_problem}. First, three detector stations of small area are considered (two could be indeed part of the same one, sitting at ground level, devoted to hard and soft muon reconstruction), all around $2 \times 2$  m$^2$ in size, following \cite{miyazaki}. Additionally, and contrary to that work, the angular information will be considered explicitly and the depth of the third (underground) station will be left as a free parameter during the optimization.

The root-mean-square error for each layer $i$ (RMSE$_i$) is introduced in order to quantify the deviations of the estimated values from the real ones, and used hereafter:
\begin{equation}
\sigma(\hat{T}_i - T_i) = \sqrt{\frac{\sum_{l=1}^{n_t}(\hat{T}_i|_l-T_i|_l)^2}{n_t}} \equiv RMSE_i \label{RMSE_eq}
\end{equation}
where $\hat{T}_i$ is the estimated (retrieved) temperature, $T_i$ the corresponding temperature data from ERA5, and the $l$ index runs in the temporal data series up to the number of measurements $n_t$. Similarly, the intrinsic time spread of the layer is defined as the standard deviation from its mean value $T_{i,0}$:
\begin{equation}
\sigma(T_i) = \sqrt{\frac{\sum_{l=1}^{n_t}(T_{i,0} -T_i|_l)^2}{n_t}}
\end{equation}

\section{Results}

\subsection{Single-channel}\label{s-c}

\begin{figure}[h!!!]
    \centering
    \includegraphics[width=\textwidth]{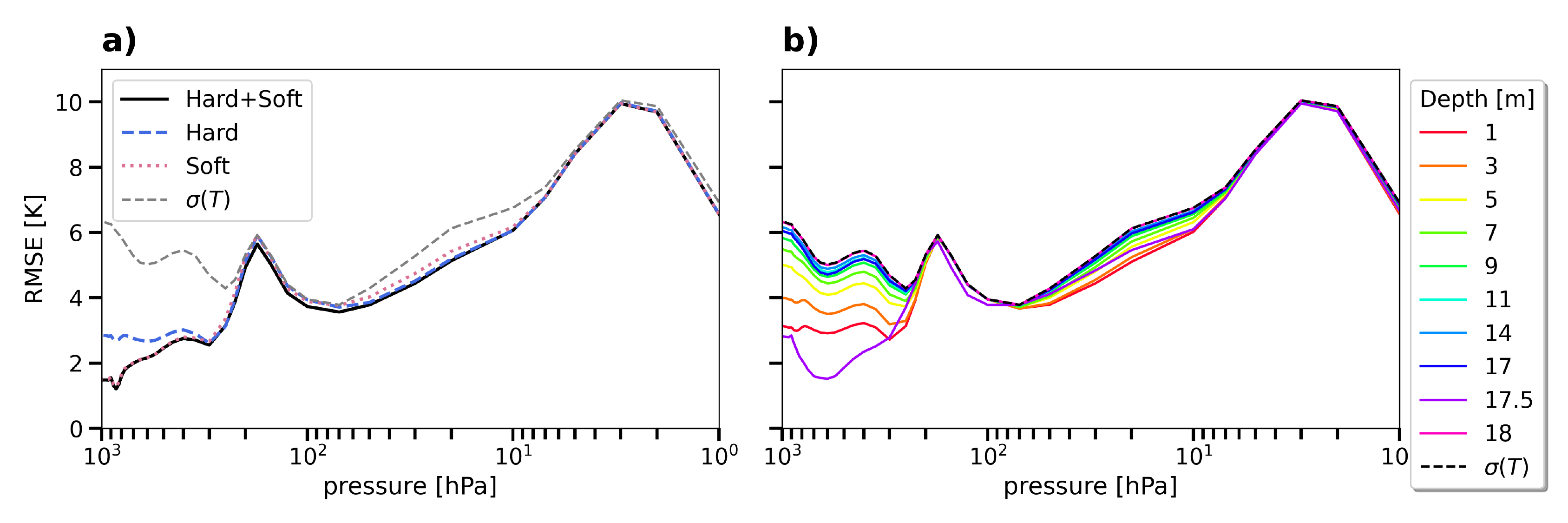}
    \caption{RMSE between estimated and real temperature, plotted as a function of the pressure level after cosmic-ray data inversion. The detector area is 2~m$ \times $2~m and the analysis has been performed in 6~h intervals. A single-channel analysis based on vertical muons has been assumed: (a) using the hard (blue line) and soft (red line) muon component, with the combination of both components shown in black; (b) using the underground muon component at different depths. The standard deviation of the temperature of each layer is overlaid in both cases (dashed grey line). The 1m-depth situation is already almost indistinguishable from the hard muon analysis shown in left. At a ``magic depth'' of 17.5~m the sensitivity to the troposphere becomes maximal and slightly improves in the stratosphere, as explained in text.}
    \label{fig:individual_singlechannel}
\end{figure}

In order to compare with previous studies, we estimate the temperature of the 37 available pressure levels using eq. \ref{eq:inverse_problem} for $j$=1, i.e., single-channel (only vertical direction, $\theta=0$-$10^\circ$) and $k=$1-3 stations. Expression \ref{eq:inverse_problem} becomes: 
\begin{equation}
    \Delta \hat{T}_i   = \sum_{k=1}^{n_{st}} c_{ki} \frac{\Delta R}{R_0}(E_{th,k},0^\circ) \bigg|_{k} 
\end{equation}
Figure \ref{fig:individual_singlechannel}a shows the RMSE for the hard (blue line) and soft (red line) components individually, together with their combination (black line). The underground component is the only one that allows the inspection of different depths (Figure \ref{fig:individual_singlechannel}b). 
The hard muon component has a threshold energy of 0.4 GeV, corresponding approximately to muons traversing 10~cm of lead. Temperature coefficients are nearly flat in this case (Figure \ref{fig:temperature_coefficients}, continuous black line), which means that rates are similarly affected by all atmospheric layers. As a consequence, it is the troposphere, the largest region of the atmosphere by weight, the one that dominates the variations of c.r. rates. Temperatures between 1000 and 300 hPa are in this way reasonably well estimated from hard muon rates (RMSE $\sim $ 3 K), however accuracy is lost in the tropopause and stratosphere (Figure \ref{fig:individual_singlechannel}a, dashed-blue line). Since the soft muon component at ground ($E<0.4$ GeV) is much more affected by the temperature of the lower layers of the atmosphere (Figure \ref{fig:temperature_coefficients}, continuous red line), it becomes more precise as a temperature estimator below 300~hPa (Figure \ref{fig:individual_singlechannel}a, dotted-red line) by suppresing the negative correlation between troposphere and tropopause regions (e.g., \cite{riadigos2020atmospheric}). The combination of both components gives a marginal improvement at this point (black line).

The temperature estimate from the inversion of underground rates is now shown in Figure \ref{fig:individual_singlechannel}b for different depths. In general, accuracy is lost with depth due to the loss of statistics as well as weights becoming closer to zero. For depths greater than 18~m the variability of the temperature estimate is indistinguishable from the temperature variability of the layer itself (dashed line in the figure). However, and similar to earlier investigations \cite{miyazaki}, even for a single-channel analysis we see some preference towards the third station being placed well underground yet at relatively shallow depths, specifically, at 17.5~m. Although this value may seem somewhat artificial, it emerges from the behaviour of the temperature coefficients at that particular depth ($E_{th} \sim 10$ GeV). It happens when the temperature coefficients in the tropopause region (100-200~hPa) reach values close to zero while still having enough weight in the troposphere to dominate the c.r. rate variations. As temperatures in the 100-150~hPa region are anticorrelated with both the troposphere and stratosphere, minimizing their contribution results in the best possible estimator for the tropospheric temperature, with a mild improvement in the stratosphere too. This delicate balance of the correlation/anti-correlation effects between troposphere, tropopause and stratosphere results in a very narrow plateau of optimum depths (10's of cm) and is highly sensitive to the weight shapes. As we will show, it achieves its full potential when angular information is considered, thereby involving a wider range of angles and thus optimal depths than seen in a single-channel analysis. Given that calculations like the ones presented here depend on weights that are theoretically estimated, a wide depth-plateau is a desirable feature to have at the outset. 

In all three analyses, a peak around the tropopause region is noticeable. Here the estimates are generally poor because the variations of temperatures in this part of the atmosphere are strongly anti-correlated with the surface variations, whose contribution to the observed rates is dominant. In addition, the temperatures of the upper stratosphere ($<50$ hPa) are also particularly difficult to capture. The reason is that this region corresponds to few percent of the total atmospheric mass. So unless the temperature coefficients would peak at that specific region and become essentially zero in the rest of the atmosphere, the effect of the remaining layers will always be dominant (i.e. they will have more weight in eq. \ref{eq:inverse_problem}). 


\begin{figure}[h!!!]
    \centering
    \includegraphics[scale=0.5]{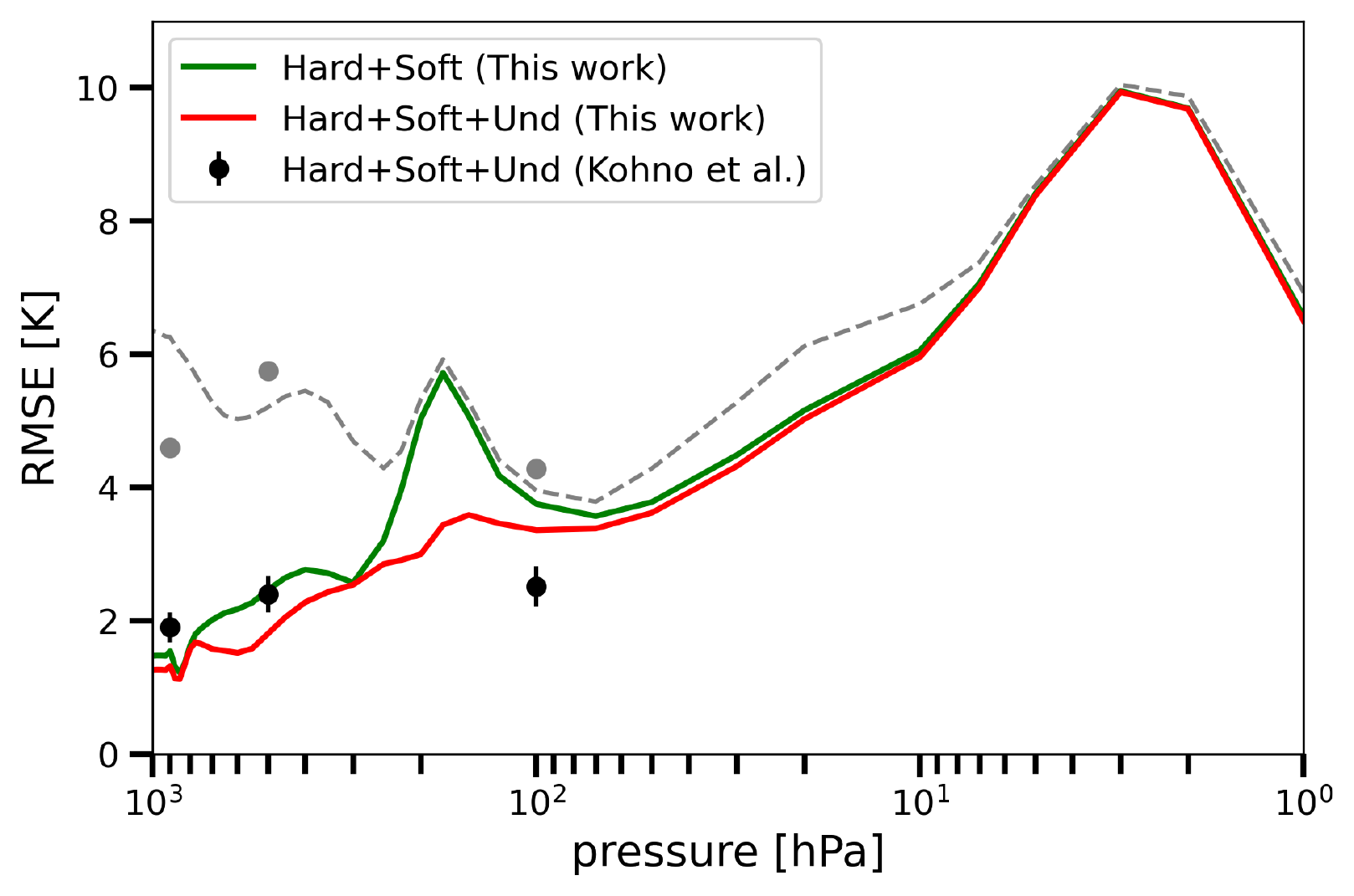}
    \caption{RMSE between estimated and real temperature, plotted as a function of the pressure level after cosmic-ray data inversion. The detector area is 2~m$ \times $2~m and the analysis has been performed in 6~h intervals. A single-channel analysis based on vertical muons has been assumed ($\theta$ below 10~deg.). The results for the combination of the hard and soft component are shown by the green line. Inclusion of a third underground station at an optimal depth around 17.5 m improves the results for layers below the 80 hPa one (red line). For comparison, the experimental results obtained around the Tokyo area in \citeA{kohno1981estimation}, also with a three-station configuration, are overlaid (black markers). The standard deviations of the temperature of each layer as seen in our data (dashed line) as well as the one on their study (grey points) are included too.}
    \label{fig:combination_singlechannel}
\end{figure}


Finally, Figure \ref{fig:combination_singlechannel} shows the simulation results for two (green) and three (red) stations, the latter for the optimal depth obtained in our study. The daily RMSE obtained from an analogous three-station (1~m$^2$) configuration in \citeA{kohno1981estimation}, with measurements performed over a 5~month period within Tokyo area, is overlaid (black points). Despite the different atmospheric conditions, the intrinsic temperature variations of the layers studied are comparable (grey points vs grey dashed line), as well as the reconstructed temperature (black points vs red line), giving support to the present analysis.

\subsection{Multi-channel}\label{m-c}

As muon stations can be easily built nowadays with a high angular resolution of the order of a degree, and given the strong dependence of atmospheric weights with zenith angle, a vast amount of additional information can be made in principle available to the inverse problem (eq. \ref{eq:inverse_problem}).
For simplicity, we chose to bin the zenith angle in 10~degree-steps (channels) ranging from 0$^\circ$ to 70$^\circ$. Clearly, there must exist more optimal ways to use the angular information, ideally keeping a reasonable statistics for each angular channel for each depth and detector area considered. Realistically, an angle-averaged $W_T$ for each channel should be used too. The straightforward binning and linear regression model chosen here aims at merely illustrating the potential (and intrinsic limits) of combining angular information with an optimal depth, hinting at which depth that is, and for which detector size.


\begin{figure}[h!!!]
    \centering
    \includegraphics[scale=0.45]{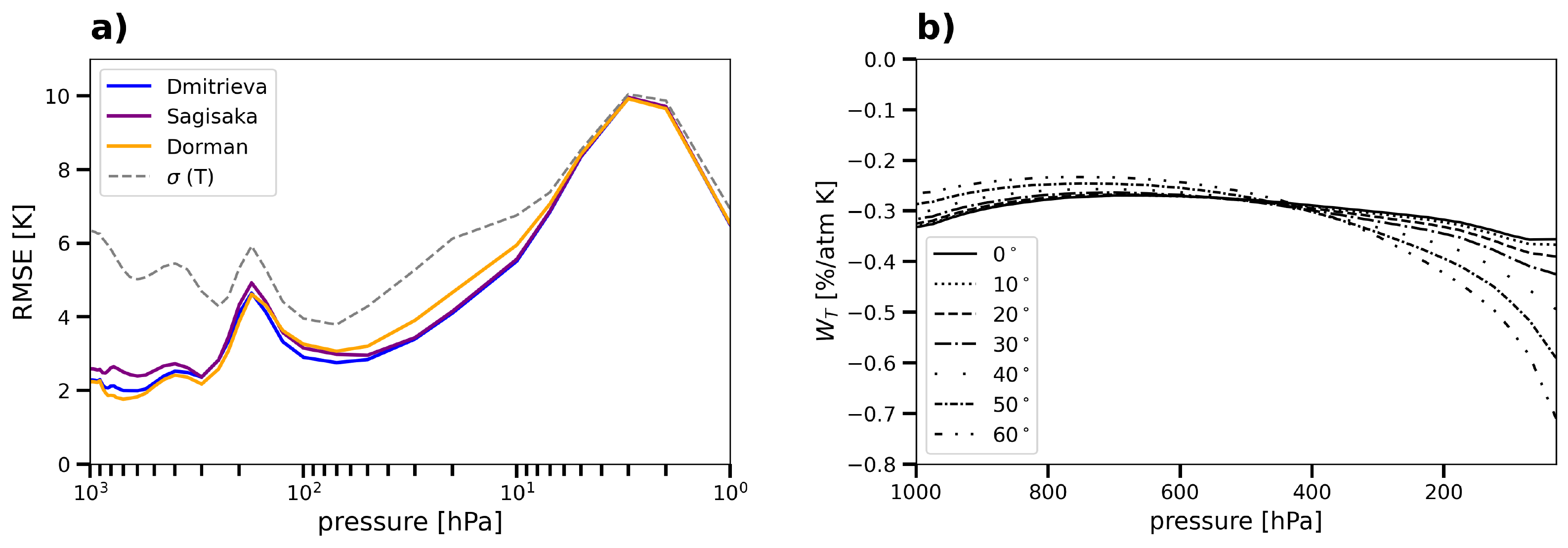}
    \caption{(a) RMSE between estimated and real temperature for each atmospheric pressure level, using the hard component. Each curve refers to the results obtained using the different databases of coefficients: \cite{sagisaka1986,dorman2004,dmitrieva2011coefficients}. (b) Distribution of temperature coefficients for hard muons, at several angles \cite{dmitrieva2011coefficients}.}
    \label{fig:hard_multchann}
\end{figure}

\begin{figure}[h!!!]
    \centering
    \includegraphics[scale=0.45]{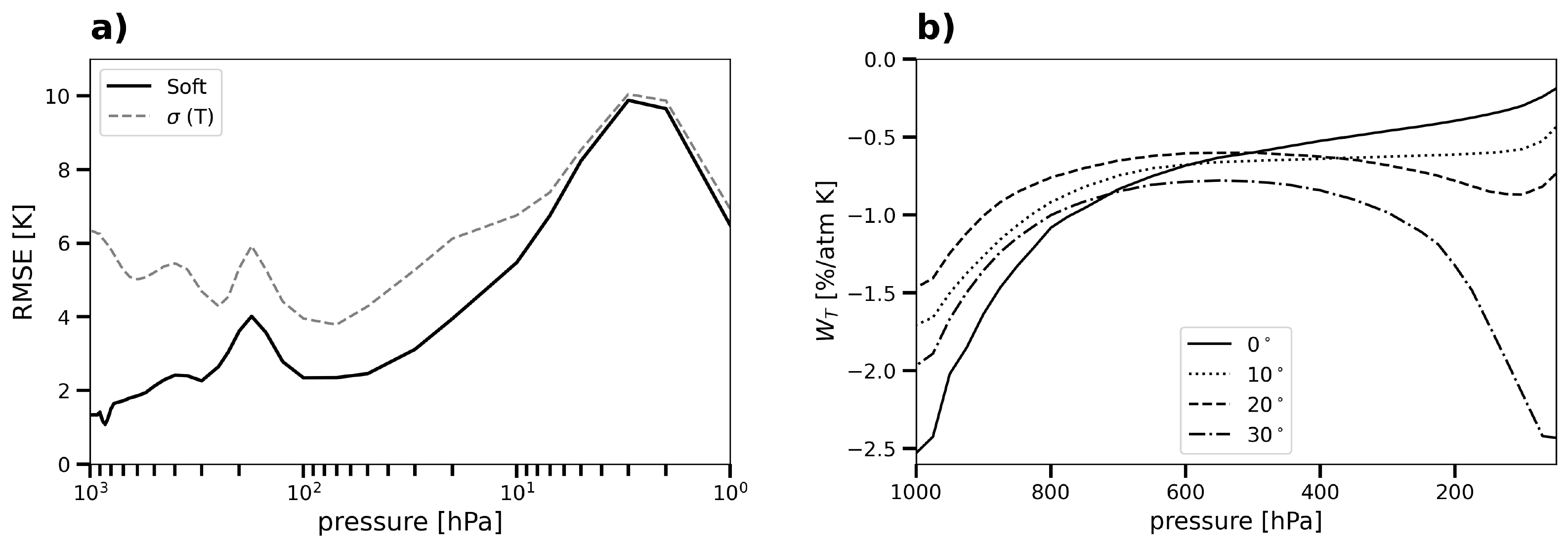}
    \caption{(a) RMSE between estimated and real temperature for each atmospheric pressure level, using the soft component. (b) Distribution of temperature coefficients for soft muons, at several angles \cite{dorman2004}.}
    \label{fig:soft_multchann}
\end{figure}

First of all, the results for muons reconstructed in the ground station (tagged as before as hard ($E>0.4$~GeV) and soft ($E<0.4$~GeV)) are shown in Figures \ref{fig:hard_multchann} and \ref{fig:soft_multchann}. As for the single-channel analysis, a typical station area of $2 \times 2$ m$^2$ has been assumed.
In the case of hard muons, Figure \ref{fig:hard_multchann}a shows the temperature predictability (RMSE, eq. \ref{RMSE_eq}) calculated employing the temperature coefficients from the three databases available to us \cite{sagisaka1986,dorman2004,dmitrieva2011coefficients}. Calculations based on Sagisaka's and Dmitrieva's coefficients are in approximate agreement, while Dorman's deviate slightly in the low stratosphere. Already at this point, the multi-channel analysis of hard muons significantly outperforms the single-channel analysis in the high atmosphere (above the level of 100 hPa) even with the latter performed through three detection stations (previous section, Figure \ref{fig:individual_singlechannel}). 
When performing the multi-channel analysis for the soft component (Figure \ref{fig:soft_multchann}a), for which only Dorman's weights seem to exist (Figure \ref{fig:soft_multchann}b), the result is even slightly better. So, with independence from the specific shape of the weights (Figure \ref{fig:hard_multchann} and \ref{fig:soft_multchann}b) the use of multiple angles within a single station represents a far better strategy when aiming at temperature retrieval in the high atmosphere than using multiple stations.

\begin{figure}
    \centering
    \includegraphics[scale=0.45]{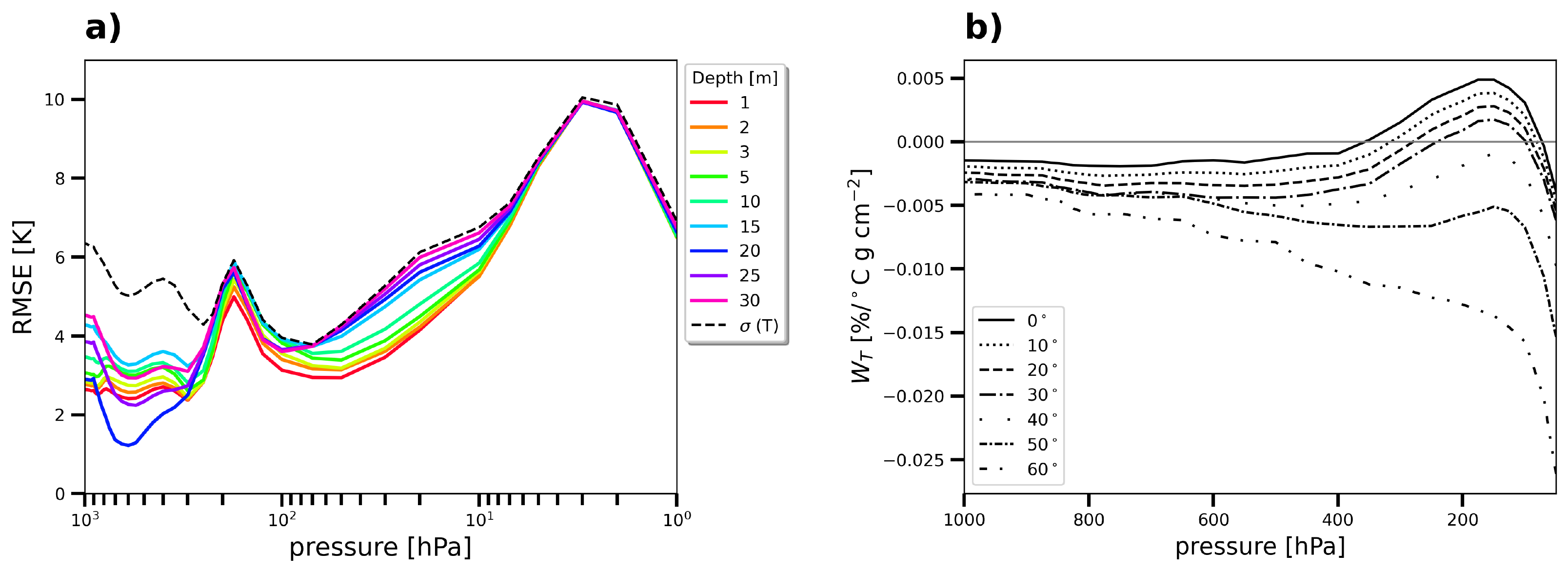}
    \caption{(a) RMSE between estimated and real temperature for each atmospheric pressure level, using underground muons. (b) Distribution of temperature coefficients for muons above 10 GeV, at several angles \cite{sagisaka1986}.}
    \label{fig:ung_multchann}
\end{figure}

\begin{figure}[h!!]
    \centering
    \includegraphics[scale=0.45]{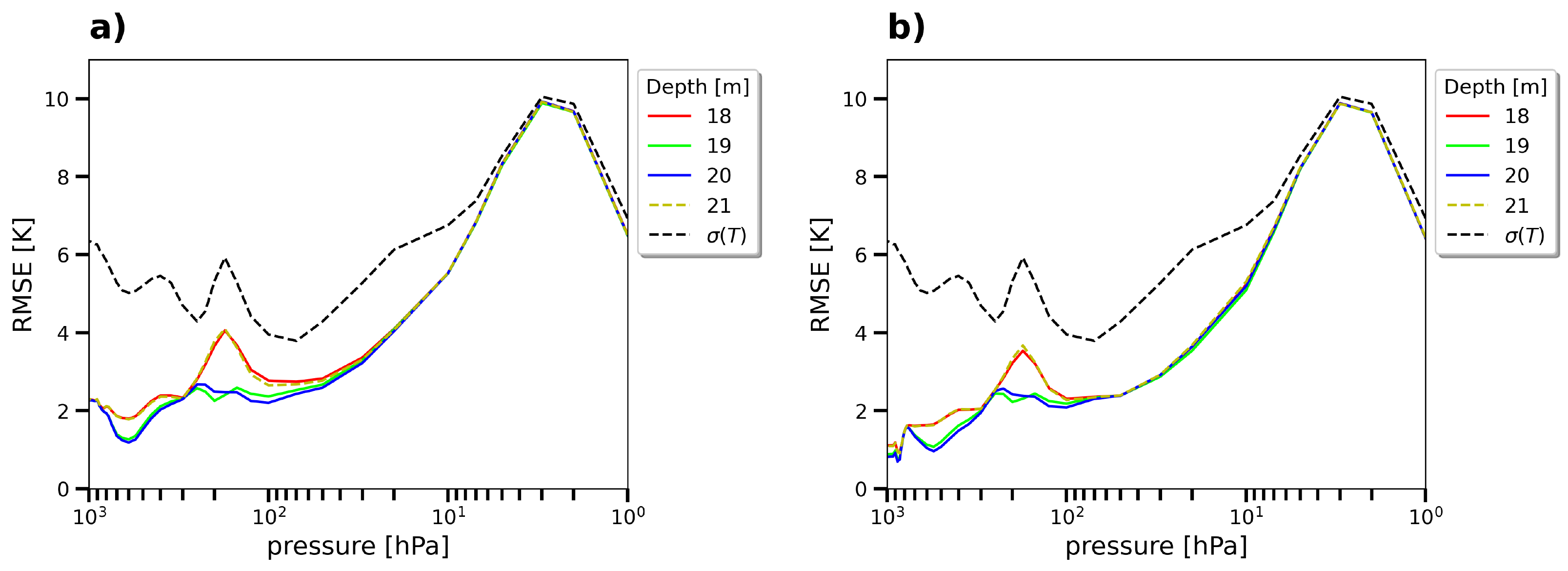}
    \caption{RMSE between estimated and real temperature for each atmospheric pressure level, in the multi-channel analysis. (a) Hard + underground muons at different depths. (b) Hard + soft + underground muons at different depths. Depths smaller than 18~m and larger than 21~m are excluded because they do not exhibit any discernible difference compared to those.}
    \label{fig:combination_multichannel}
\end{figure}

Figure \ref{fig:ung_multchann}a shows now the RMSE for underground stations placed at different depths, assuming a uniform-thickness soil slab. As in the single-channel analysis, statistics limits the station capabilities quickly as a function of depth, and so for 30-m depth a $2\times2$m$^2$ detector is already insensitive to temperatures above the tropopause (magenta line). As the distribution gets more peaked at small zenith angles the deeper the station is placed, the usefulness of multiple angular bins becomes more limited and the best global description is obtained again (as in the single-channel analysis) for near-surface detectors (1~m deep, red line). Indeed, the RMSE for a 1~m-deep station is very close the one obtained in Figure \ref{fig:hard_multchann} for hard muons. Interestingly, however, a 20~m-deep station would perform significantly better for tropospheric levels below 300 hPa. This reproduces the effect observed in the single-channel analysis, although for a broader range of depths. It becomes even more apparent when combining several stations (Figure \ref{fig:combination_multichannel}): weight coefficients in the range 19-20~m, despite showing a low sensitivity to temperature (up to $\times$~100 less than weights at ground) exhibit a strong dependence with angle and atmospheric height, going from positive to negative values in the tropopause region. This can be used beneficially in the regression to minimize the correlation/anti-correlation effects in the troposphere-tropopause-stratosphere regions. At the same time, the relatively shallow depth of the station is compatible with a moderate statistical noise in counting, even for angles far from the vertical.

The combined analysis of the different muon components/stations is shown in Figure \ref{fig:combination_multichannel} for the following two cases: hard + underground (Figure \ref{fig:combination_multichannel}a) and hard + soft + underground (Figure \ref{fig:combination_multichannel}b). While the inclusion of soft muons brings a sizeable improvement in the troposphere region, it does not drive the fact that an optimum depth exists.

\begin{figure} [h!!!]
    \centering
    \includegraphics[scale=0.6]{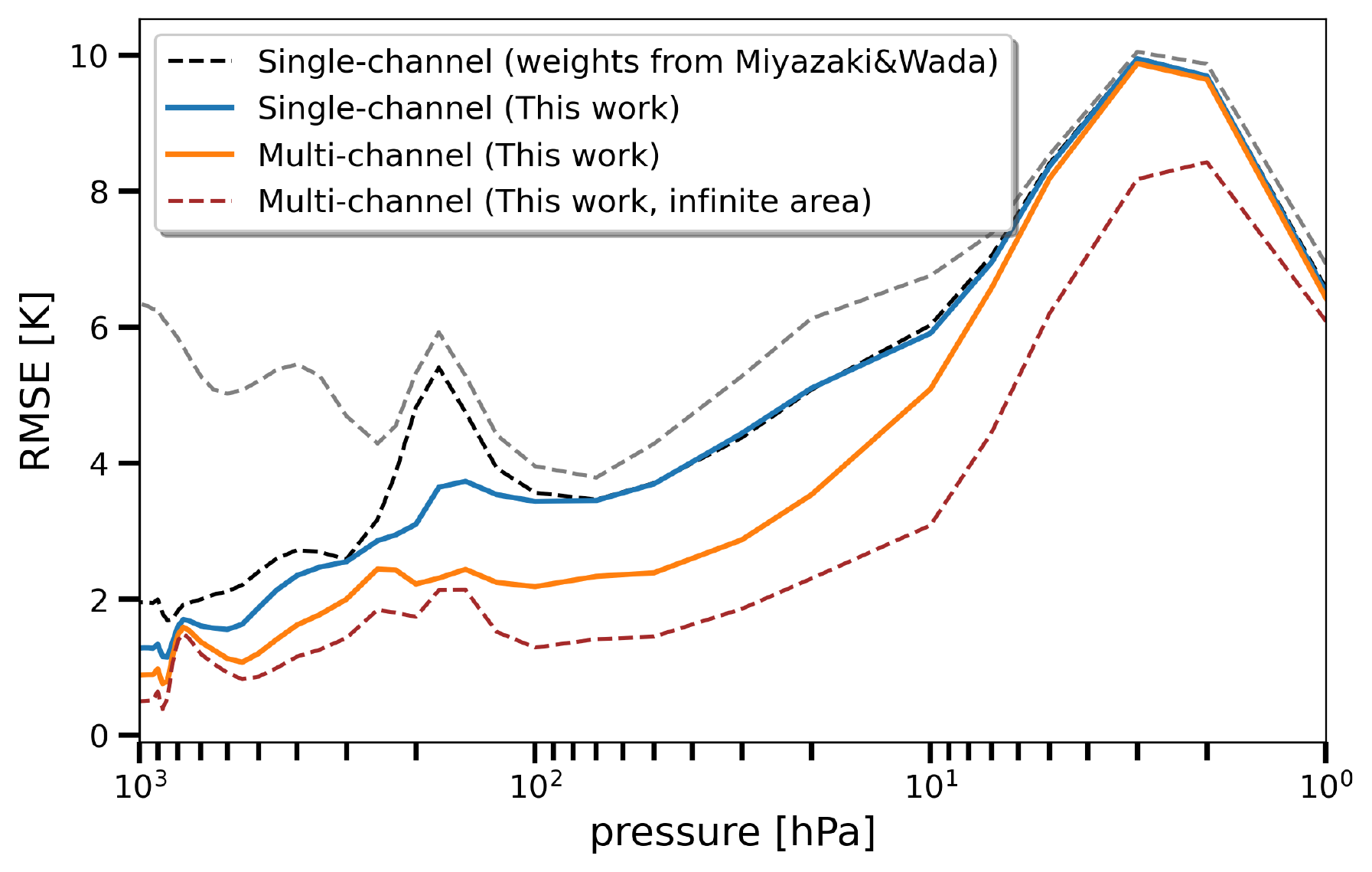}
    \caption{RMSE between estimated and real temperature for each atmospheric pressure level. Observations performed in 6~h time intervals and for 2~m $\times$ 2~m detection area. Black dashed:  three-station/single-channel analysis using weights from \cite{miyazaki}. Blue: three-station/single-channel analysis at an optimized underground depth. Orange: three-station/multi-channel analysis at optimized underground depth. Dashed red: three-station/multi-channel analysis at optimized underground depth, neglecting statistical fluctuations in particle counting. The grey dashed line represents the intrinsic spread of the temperatures for each atmospheric layer.}
    \label{fig:final_results}
\end{figure}

At first glance, the optimum depth derived from this analysis would seem to be in agreement with that chosen in the three-station/single-channel simulation of \cite{miyazaki}. In their work they selected a depth of 55 m.w.e, which is equivalent to a threshold energy of $\sim$11~GeV or, in other words, to a soil-thickness of 20~m. However, the early temperature coefficients there assumed are much more peaked than the ones used here, and would correspond to a threshold of 50~GeV (80 m-depth) if resorting to more modern estimates as those shown in Figure \ref{fig:temperature_coefficients}. The performance of the two methods is very different too, as can be appreciated in Figure \ref{fig:final_results} where the RMSE from the three-station/single-channel analysis as proposed in Miyazaki and Wada (black dashed line) is compared with the present one (blue line). Despite the assumed depth is the same in both cases, the difference in the assumed weights makes all the difference, allowing us to establish the relevance of the 20~m-depth as an actual optimum for atmospheric studies. The improvement is even more apparent when considering a multi-channel analysis (orange line): up to a factor of two or more can be gained in critical atmospheric regions like the tropopause and stratosphere compared to earlier simulation work. The reconstruction reaches a best value of 0.8~K at 850~hPa and a worse one around 2.2~K for the tropopause region and up to 50~hPa, becoming the temperature intrinsically inaccessible above the 10~hPa layer. Moreover, Miyazaki and Wada estimated RMSE-values between 1 and 3~K when disregarding statistical noise from counting, for seven pressure levels between 1000 and 100~hPa. Interpreting that as the intrinsic limit to the inversion problem and comparing to the analogous result in our analysis (red-dashed line in Figure \ref{fig:final_results}), it can be concluded that a three-station/multi-channel analysis with optimized depth provides an overall improvement of around a factor 2 also in that situation.

\section{Discussion}

Within the relatively simple inversion algorithm proposed in this work, the width of the optimal-depth plateau in Figure \ref{fig:combination_multichannel} exhibits a non-trivial dependence with the chosen angular binning. For depths around the optimal one, this is exacerbated since the overall correlation between rates and temperature (eq. S1) changes sign (hence, it vanishes) as a function of the angular channel. It does so in a way that is both abrupt and critically dependent on the precise shapes of the weights. While this fact suggests that a finer resolution is desirable (1 degree is technically possible without great effort), we opted to leave such a study outside this work. The reason is two-fold: i) the binning becomes too thin compared to the four angular bins available for our underground coefficients in \cite{sagisaka1986}, and so, in the absence of new calculations of those, the present simulation work would depend largely on the interpolation method and ii) the $\times 10$ increase in the number of fitting parameters would require of a more dedicated optimization study than intended here. In our case we have relied on the python package $Statsmodels$, and the Ordinary Least Squares method included in it, without constraints in the fitting parameters (some examples of regression plots can be found in Figure S5 of Supplementary Information). The results were little sensitive to the method chosen or how the regression was conditioned (initial values, parameter range, function tolerance and linear constraints between variables). Studies performed with a mildly increased binning (5 degrees) show indeed that the results of the regression within the 19-20~m plateau become much more stable.

Once the existence of an optimal depth-plateau has been established, it is important to understand how the performance of the inversion algorithm depends on the size of the detection area and the presence of systematic errors in particle counting (that we have simulated getting random samples from a Gaussian distribution with the width being a certain percentage of the average rate and subsequently adding them to the data series). To make the latter more realistic, we follow the typical procedure of adding extra noise to the data in different levels (low, medium, and high) to analyze the performance of the technique. It was observed that larger areas present just marginal gains compared to a 4 m$^2$ detector and, at the same time, the performance of the proposed configuration deteriorates severely above 0.3\% systematic variations (see Figures S2 and S3 included in the Supplementary Information). This poses a very stringent requirement for the detection system, whose overall efficiency should be kept stable within these values. From this viewpoint, plastic detectors coupled to photon sensors represent a most natural choice, although a gaseous detector with redundant layers could become more affordable/practical at the expense of a larger design complexity.

\begin{figure}[h!!!]
    \centering
    \includegraphics[scale=0.6]{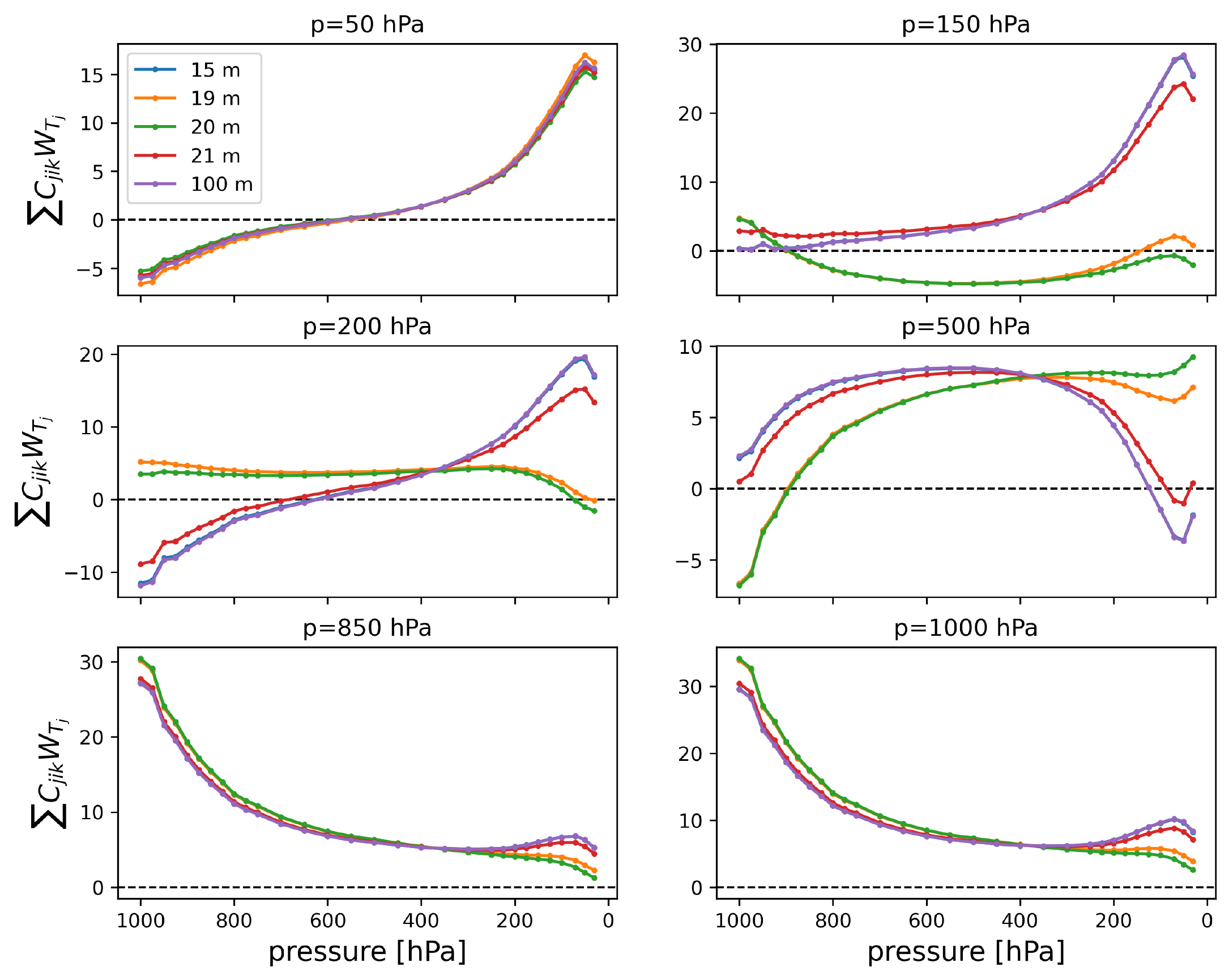}
    \caption{Combined temperature coefficients calculated for the atmospheric pressure levels of 50, 150, 200, 500, 850 and 1000 hPa in the three-station/multi-channel configuration. The curves represent different depths of the underground station: 15, 19, 20, 21 and 100 m.}
    \label{fig_apend:comb_coeff}
\end{figure}

Results presented here are difficult to interpret from an atmospheric physics perspective, so in order to get a better grasp of how the minimization process works, a close examination of what we define here as ``combined temperature coefficients" will show to be useful. For that we rewrite equation \ref{eq:inverse_problem} taking into account expression \ref{eq:temperature_effect}:

\begin{equation}
    \Delta \hat{T}_i = \sum_{k=1}^{n_{st}} \sum_{j=1}^{n_{ch}} c_{jki} \left( \sum_{p=1}^{n_l} W_T(E_{th,jk},\theta_j,h_p) \Delta T_p \Delta h_p \right)\bigg|_{k} 
\end{equation}
Rearranging the terms for the same pressure level $p$ we obtain:
\begin{equation}
    \Delta \hat{T}_i = \sum_{p=1}^{n_l} \left(\sum_{k=1}^{n_{st}} \sum_{j=1}^{n_{ch}} c_{jki} W_T(E_{th,jk},\theta_j,h_p) \right) \Delta T_p \Delta h_p 
\label{eq:combined_coeff}
\end{equation}
from where we define the combined temperature coefficient as:
\begin{equation}
    W_{pi} \equiv \sum_{k=1}^{n_{st}} \sum_{j=1}^{n_{ch}} c_{jki} W_T(E_{th,jk},\theta_j,h_p)
\end{equation}
When solving the inverse problem for the temperature of a certain pressure layer, one may expect that the coefficients $c_{jki}$ of the regression should in principle have values such that in eq. \ref{eq:combined_coeff} the combined coefficients are able to enhance the $p$-term of the same temperature layer ($i$), minimizing the contribution from the rest. As the shapes of the temperature coefficients are not flexible enough to accommodate this condition for any arbitrary layer, the correlation between atmospheric layers becomes an essential ingredient. Figure \ref{fig_apend:comb_coeff} shows some examples of the combined coefficients obtained for the retrieval of the temperature at 50, 150, 200, 500, 850, and 1000~hPa for different depths of interest: 15, 19, 20, 21, and 100~m. Further information can be found in the Supplementary Information. 

\begin{figure}[h!!!]
    \centering
    \includegraphics[width=\textwidth]{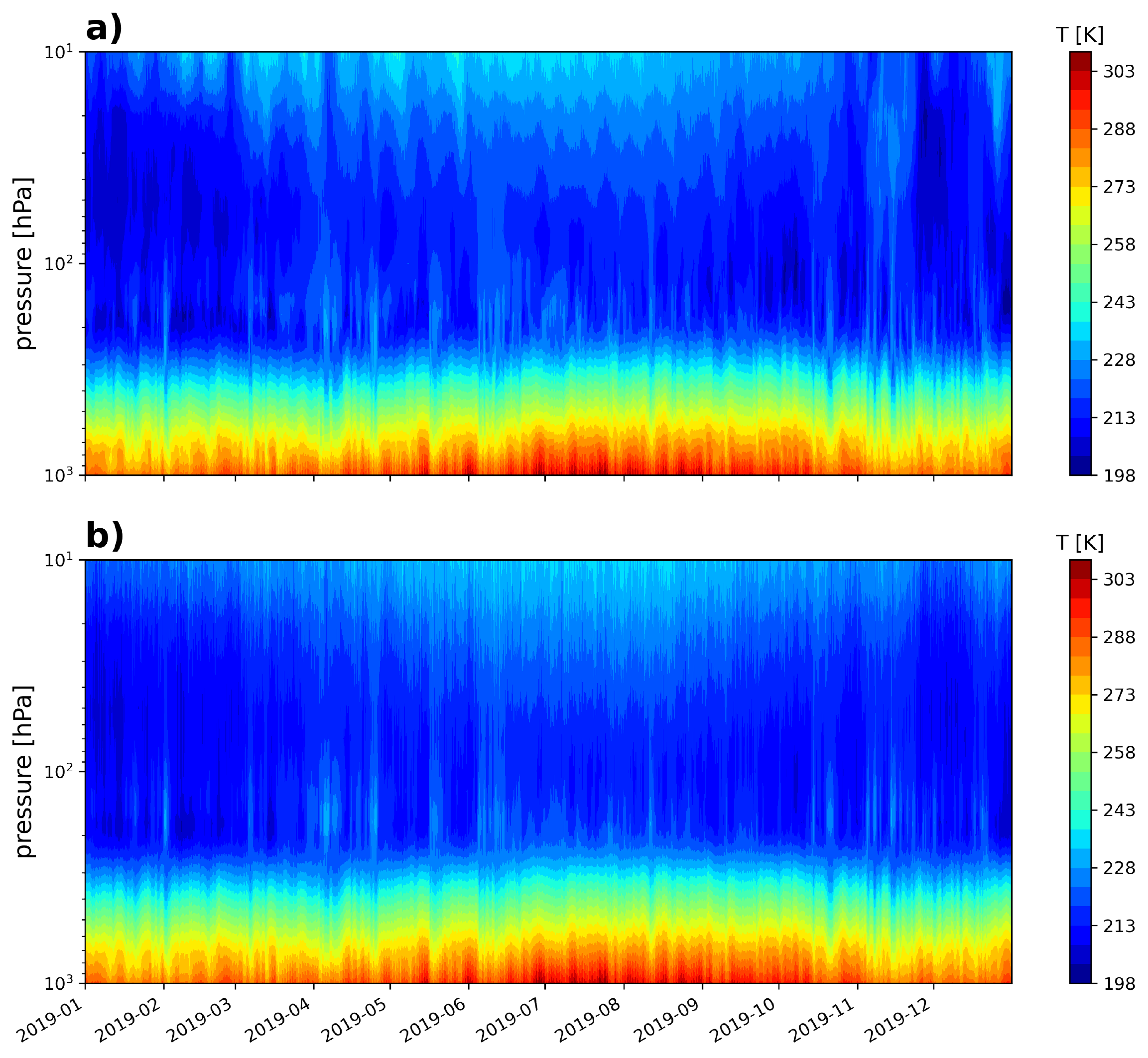}
    \caption{(a) Observed temperature for 2019. (b) Estimated temperature for 2019 using the detector configuration and data analysis discussed in text.}
    \label{fig:estimations}
\end{figure}

\begin{figure}[h!!!]
    \centering
    \includegraphics[width=\textwidth]{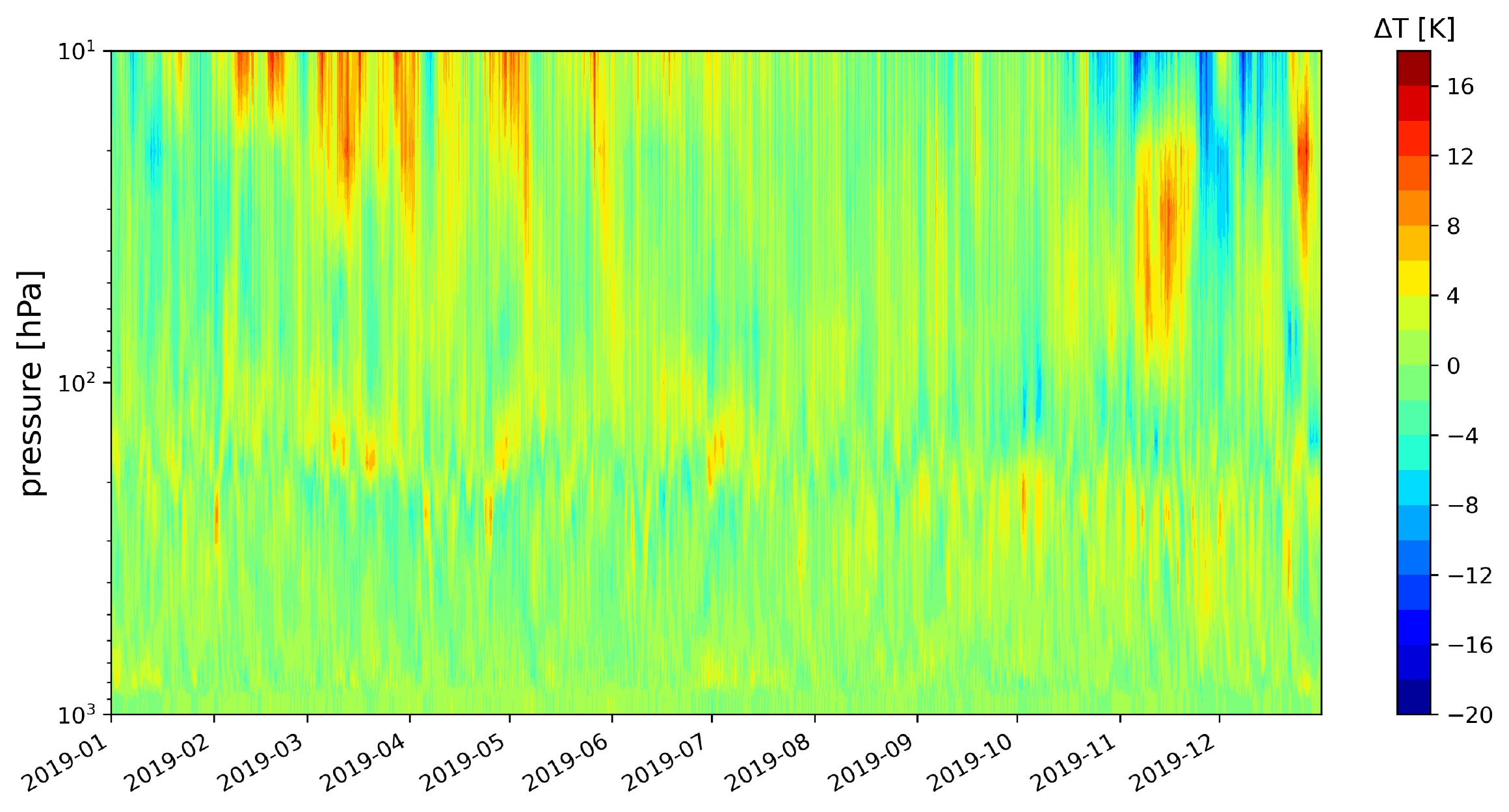}
    \caption{Difference between observed and estimated temperature for 2019 using the detector configuration and data analysis discussed in the text.}
    \label{fig:diff_temps}
\end{figure}

Overall, our study shows that the introduction of different angles into the analysis of the inverse problem helps improving the estimates of the atmospheric temperature profile dramatically. A temperature predictability (RMSE) ranging from 0.8-1K in the low troposphere to 2.2~K in the 50~hPa level was obtained, deteriorating for higher stratospheric layers. For a temperature monitoring emplacement dedicated to improve atmospheric forecasts, this range of heights is more than enough since the most interesting atmospheric phenomena occur in the tropospheric layer. However, there is a unique phenomenon that takes place in the stratosphere and is attracting a lot of atmospheric scientists due to its capacity to modify the weather at the surface. We are referring to Sudden Stratospheric Warmings (SSW). A SSW is an event that occurs in polar vortices when the stratospheric temperature suffers an abrupt increase in a short period of time. In such events, the vortex may collapse, releasing cold air towards lower latitudes that could impact the surface weather. This situation is more likely to happen in the northern hemisphere. In many cases, the monitoring of the temperature at 10~hPa (for which a modest 5~K RMSE was obtained in this analysis) is useful when a major event occurs. Still, the observation of any other lower stratospheric level provides very valuable information about these events. In such a case, the setup proposed in this work would be at least complementary to this kind of research.  For illustration, a comparison between the input temperatures and the ones estimated with the inversion method proposed in text is shown in Figure \ref{fig:estimations}. The difference between the observed and estimated temperatures is also included in Figure \ref{fig:diff_temps}. The BIAS calculated for these predictions is zero for the pressure levels presented. However, it can be seen that the upper atmosphere displays the most significant errors, which correspond to the occurrence of the aforementioned SSW events. The model loses detail when capturing these events, as we have mentioned above, yet it is able to reproduce its influence on lower levels (between 20 and 200~hPa).

\section{Conclusions}

The purpose of the present study was to determine the best configuration of a monitoring station integrated by cosmic-ray telescopes. One of the most significant findings to emerge from this study is the possibility to retrieve the temperature of the atmosphere from the surface with good accuracy up to a considerable height ($\sim$ 20 km). An implication of this is that atmospheric muon detectors can be used in scientific research beyond the field of astrophysics.

Returning to the question posed at the beginning of this paper, it is now possible to state that multi-directional telescopes can enhance the estimate of atmospheric temperatures. Our results have shown that one can achieve a high degree of accuracy with error margins between 0.8 and 2.2 K that could be improved in practice applying more advanced statistical techniques such as those employed in satellite observations \cite{eyre1989inversion,rodgers2000inverse}.  
For 4~m$^2$-scale detectors, this performance requires of an outstanding detector stability (below 0.3\% on its counting efficiency). While detector inefficiencies to minimum ionizing particles down to 0.1\% are not alien to particle physics instrumentation, the requirement will pose significant constraints on the chosen technology and detector design.

On the other hand, the current work was limited by the use of simulated c.r. data and further work needs to be done in the area of experimentation to evaluate the actual limits of the estimates. In spite of this, our findings establish several courses of action for future research. A good line for future work would be to contrast the retrieved temperatures from real cosmic-ray data against temperature data from balloon measurements. Additionally, we have found evidence of an optimal depth to place one of the detectors. Depths of 19-20 m seem feasible and affordable, with no need to go deep into a mountain as is done for other c.r. research. Such an underground station could be easily located in dams, tunnels or subway stations for instance. Moreover, continuous measurements of vertical temperature provide in this way would no doubt complement satellite measurements, as this technology would be much more affordable and easy to assemble and maintain.

\acknowledgments
We want to thank J.A. Garzón-Heydt and D. García-Castro (IGFAE) for stimulating discussions on the atmospheric effect and M. Slupecki, W. H. Trzaska (Univ. of Jyväskylä) for explanations on underground detector data.

We gratefully acknowledge financial support by Xunta de Galicia under Research Grant No. 2021-PG036-1. We also thank financial support from Xunta de Galicia (Centro singular de investigación de Galicia accreditation 2019-2022), by European Union ERDF, and by the ``María  de Maeztu''  Units  of  Excellence program  MDM-2016-0692 and the Spanish Research State Agency. D.G.D. acknowledges the Ramon y Cajal program (Spain) under contract number RYC-2015-18820.

Authors are thankful to ECMWF for providing the ERA5 reanalysis. This data is cited in the references and can be found in the following repository from the Copernicus Climate Change Service (C3S): https://doi.org/10.24381/cds.bd0915c6


%
%

\bibliography{bibliography}

\begin{thebibliography}{}

\bibitem [\protect \citeauthoryear {%
Adamson%
\ \protect \BOthers {.}}{%
Adamson%
\ \protect \BOthers {.}}{%
{\protect \APACyear {2010}}%
}]{%
adamson2010minos}
\APACinsertmetastar {%
adamson2010minos}%
\begin{APACrefauthors}%
Adamson, P.%
, Andreopoulos, C.%
, Arms, K.%
, Armstrong, R.%
, Auty, D.%
, Ayres, D.%
\BDBL {}others%
\end{APACrefauthors}%
\unskip\
\newblock
\APACrefYearMonthDay{2010}{}{}.
\newblock
{\BBOQ}\APACrefatitle {Observation of muon intensity variations by season with
  the {MINOS} far detector} {Observation of muon intensity variations by season
  with the {MINOS} far detector}.{\BBCQ}
\newblock
\APACjournalVolNumPages{Physical Review D}{81}{1}{012001}.
\PrintBackRefs{\CurrentBib}

\bibitem [\protect \citeauthoryear {%
Aglietta%
\ \protect \BOthers {.}}{%
Aglietta%
\ \protect \BOthers {.}}{%
{\protect \APACyear {1998}}%
}]{%
aglietta1998depthintensity}
\APACinsertmetastar {%
aglietta1998depthintensity}%
\begin{APACrefauthors}%
Aglietta, M.%
, Alpat, B.%
, Alyea, E.%
, Antonioli, P.%
, Badino, G.%
, Bari, G.%
\BDBL {}others%
\end{APACrefauthors}%
\unskip\
\newblock
\APACrefYearMonthDay{1998}{}{}.
\newblock
{\BBOQ}\APACrefatitle {Muon “depth-intensity” relation measured by the
  {LVD} underground experiment and cosmic-ray muon spectrum at sea level} {Muon
  “depth-intensity” relation measured by the {LVD} underground experiment
  and cosmic-ray muon spectrum at sea level}.{\BBCQ}
\newblock
\APACjournalVolNumPages{Physical Review D}{58}{9}{092005}.
\PrintBackRefs{\CurrentBib}

\bibitem [\protect \citeauthoryear {%
Ambrosio%
\ \protect \BOthers {.}}{%
Ambrosio%
\ \protect \BOthers {.}}{%
{\protect \APACyear {1997}}%
}]{%
ambrosio1997macro}
\APACinsertmetastar {%
ambrosio1997macro}%
\begin{APACrefauthors}%
Ambrosio, M.%
, Antolini, R.%
, Auriemma, G.%
, Baker, R.%
, Baldini, A.%
, Barbarino, G.%
\BDBL {}others%
\end{APACrefauthors}%
\unskip\
\newblock
\APACrefYearMonthDay{1997}{}{}.
\newblock
{\BBOQ}\APACrefatitle {Seasonal variations in the underground muon intensity as
  seen by {MACRO}} {Seasonal variations in the underground muon intensity as
  seen by {MACRO}}.{\BBCQ}
\newblock
\APACjournalVolNumPages{Astroparticle Physics}{7}{1-2}{109--124}.
\PrintBackRefs{\CurrentBib}

\bibitem [\protect \citeauthoryear {%
Baesso%
, Cussans%
, Thomay%
\BCBL {}\ \BBA {} Velthuis%
}{%
Baesso%
\ \protect \BOthers {.}}{%
{\protect \APACyear {2014}}%
}]{%
baesso2014toward}
\APACinsertmetastar {%
baesso2014toward}%
\begin{APACrefauthors}%
Baesso, P.%
, Cussans, D.%
, Thomay, C.%
\BCBL {}\ \BBA {} Velthuis, J.%
\end{APACrefauthors}%
\unskip\
\newblock
\APACrefYearMonthDay{2014}{}{}.
\newblock
{\BBOQ}\APACrefatitle {Toward a {RPC}-based muon tomography system for cargo
  containers.} {Toward a {RPC}-based muon tomography system for cargo
  containers.}{\BBCQ}
\newblock
\APACjournalVolNumPages{Journal of Instrumentation}{9}{10}{C10041}.
\PrintBackRefs{\CurrentBib}

\bibitem [\protect \citeauthoryear {%
Barrett%
, Bollinger%
, Cocconi%
, Eisenberg%
\BCBL {}\ \BBA {} Greisen%
}{%
Barrett%
\ \protect \BOthers {.}}{%
{\protect \APACyear {1952}}%
}]{%
barrett1952effecttemp}
\APACinsertmetastar {%
barrett1952effecttemp}%
\begin{APACrefauthors}%
Barrett, P\BPBI H.%
, Bollinger, L\BPBI M.%
, Cocconi, G.%
, Eisenberg, Y.%
\BCBL {}\ \BBA {} Greisen, K.%
\end{APACrefauthors}%
\unskip\
\newblock
\APACrefYearMonthDay{1952}{}{}.
\newblock
{\BBOQ}\APACrefatitle {Interpretation of cosmic-ray measurements far
  underground} {Interpretation of cosmic-ray measurements far
  underground}.{\BBCQ}
\newblock
\APACjournalVolNumPages{Reviews of Modern Physics}{24}{3}{133}.
\PrintBackRefs{\CurrentBib}

\bibitem [\protect \citeauthoryear {%
Blanco%
\ \protect \BOthers {.}}{%
Blanco%
\ \protect \BOthers {.}}{%
{\protect \APACyear {2020}}%
}]{%
blanco2020ship}
\APACinsertmetastar {%
blanco2020ship}%
\begin{APACrefauthors}%
Blanco, A.%
, Clemencio, F.%
, Fonte, P.%
, Franco, C.%
, Leonardo, N.%
, Lopes, L.%
\BDBL {}Soares, G.%
\end{APACrefauthors}%
\unskip\
\newblock
\APACrefYearMonthDay{2020}{}{}.
\newblock
{\BBOQ}\APACrefatitle {The {SHiP} timing detector based on {MRPC}} {The {SHiP}
  timing detector based on {MRPC}}.{\BBCQ}
\newblock
\APACjournalVolNumPages{Journal of Instrumentation}{15}{10}{C10017}.
\PrintBackRefs{\CurrentBib}

\bibitem [\protect \citeauthoryear {%
Bogdanova%
, Gavrilov%
, Kornoukhov%
\BCBL {}\ \BBA {} Starostin%
}{%
Bogdanova%
\ \protect \BOthers {.}}{%
{\protect \APACyear {2006}}%
}]{%
bogdanova2006cosmic}
\APACinsertmetastar {%
bogdanova2006cosmic}%
\begin{APACrefauthors}%
Bogdanova, L.%
, Gavrilov, M.%
, Kornoukhov, V.%
\BCBL {}\ \BBA {} Starostin, A.%
\end{APACrefauthors}%
\unskip\
\newblock
\APACrefYearMonthDay{2006}{}{}.
\newblock
{\BBOQ}\APACrefatitle {Cosmic muon flux at shallow depths underground} {Cosmic
  muon flux at shallow depths underground}.{\BBCQ}
\newblock
\APACjournalVolNumPages{Physics of Atomic Nuclei}{69}{8}{1293--1298}.
\PrintBackRefs{\CurrentBib}

\bibitem [\protect \citeauthoryear {%
Dmitrieva%
, Kokoulin%
, Petrukhin%
\BCBL {}\ \BBA {} Timashkov%
}{%
Dmitrieva%
\ \protect \BOthers {.}}{%
{\protect \APACyear {2011}}%
}]{%
dmitrieva2011coefficients}
\APACinsertmetastar {%
dmitrieva2011coefficients}%
\begin{APACrefauthors}%
Dmitrieva, A.%
, Kokoulin, R.%
, Petrukhin, A.%
\BCBL {}\ \BBA {} Timashkov, D.%
\end{APACrefauthors}%
\unskip\
\newblock
\APACrefYearMonthDay{2011}{}{}.
\newblock
{\BBOQ}\APACrefatitle {Corrections for temperature effect for ground-based muon
  hodoscopes} {Corrections for temperature effect for ground-based muon
  hodoscopes}.{\BBCQ}
\newblock
\APACjournalVolNumPages{Astroparticle Physics}{34}{6}{401--411}.
\PrintBackRefs{\CurrentBib}

\bibitem [\protect \citeauthoryear {%
L.~Dorman%
}{%
L.~Dorman%
}{%
{\protect \APACyear {1972}}%
}]{%
dorman1972}
\APACinsertmetastar {%
dorman1972}%
\begin{APACrefauthors}%
Dorman, L.%
\end{APACrefauthors}%
\unskip\
\newblock
\APACrefYearMonthDay{1972}{}{}.
\newblock
{\BBOQ}\APACrefatitle {The meteorological effects of cosmic rays} {The
  meteorological effects of cosmic rays}.{\BBCQ}
\newblock
\APACjournalVolNumPages{Nauka press, Russia}{}{}{}.
\PrintBackRefs{\CurrentBib}

\bibitem [\protect \citeauthoryear {%
L\BPBI I.~Dorman%
}{%
L\BPBI I.~Dorman%
}{%
{\protect \APACyear {2004}}%
}]{%
dorman2004}
\APACinsertmetastar {%
dorman2004}%
\begin{APACrefauthors}%
Dorman, L\BPBI I.%
\end{APACrefauthors}%
\unskip\
\newblock
\APACrefYear{2004}.
\newblock
\APACrefbtitle {Cosmic rays in the {Earth’s} atmosphere and underground}
  {Cosmic rays in the {Earth’s} atmosphere and underground}\ (\BVOL~303).
\newblock
\APACaddressPublisher{}{Springer Science \& Business Media}.
\PrintBackRefs{\CurrentBib}

\bibitem [\protect \citeauthoryear {%
Duperier%
}{%
Duperier%
}{%
{\protect \APACyear {1949}}%
}]{%
duperier1949positive}
\APACinsertmetastar {%
duperier1949positive}%
\begin{APACrefauthors}%
Duperier, A.%
\end{APACrefauthors}%
\unskip\
\newblock
\APACrefYearMonthDay{1949}{}{}.
\newblock
{\BBOQ}\APACrefatitle {The Meson Intensity at the Surface of the {Earth} and
  the Temperature at the Production Level} {The meson intensity at the surface
  of the {Earth} and the temperature at the production level}.{\BBCQ}
\newblock
\APACjournalVolNumPages{Proceedings of the Physical Society. Section
  A}{62}{11}{684}.
\PrintBackRefs{\CurrentBib}

\bibitem [\protect \citeauthoryear {%
Eyre%
}{%
Eyre%
}{%
{\protect \APACyear {1989}}%
}]{%
eyre1989inversion}
\APACinsertmetastar {%
eyre1989inversion}%
\begin{APACrefauthors}%
Eyre, J.%
\end{APACrefauthors}%
\unskip\
\newblock
\APACrefYearMonthDay{1989}{}{}.
\newblock
{\BBOQ}\APACrefatitle {Inversion of cloudy satellite sounding radiances by
  nonlinear optimal estimation. I: Theory and simulation for {TOVS}} {Inversion
  of cloudy satellite sounding radiances by nonlinear optimal estimation. i:
  Theory and simulation for {TOVS}}.{\BBCQ}
\newblock
\APACjournalVolNumPages{Quarterly Journal of the Royal Meteorological
  Society}{115}{489}{1001--1026}.
\PrintBackRefs{\CurrentBib}

\bibitem [\protect \citeauthoryear {%
Eyre%
}{%
Eyre%
}{%
{\protect \APACyear {1991}}%
}]{%
eyre1991satellites}
\APACinsertmetastar {%
eyre1991satellites}%
\begin{APACrefauthors}%
Eyre, J.%
\end{APACrefauthors}%
\unskip\
\newblock
\APACrefYearMonthDay{1991}{}{}.
\newblock
{\BBOQ}\APACrefatitle {Inversion methods for satellite sounding data}
  {Inversion methods for satellite sounding data}.{\BBCQ}
\newblock
\APACjournalVolNumPages{ECMWF Meteorological Training Course Lecture
  Series}{}{}{}.
\PrintBackRefs{\CurrentBib}

\bibitem [\protect \citeauthoryear {%
Gaisser%
}{%
Gaisser%
}{%
{\protect \APACyear {1991}}%
}]{%
gaisser1991}
\APACinsertmetastar {%
gaisser1991}%
\begin{APACrefauthors}%
Gaisser, T\BPBI K.%
\end{APACrefauthors}%
\unskip\
\newblock
\APACrefYearMonthDay{1991}{}{}.
\newblock
{\BBOQ}\APACrefatitle {Cosmic Rays and Particle Physics} {Cosmic rays and
  particle physics}.{\BBCQ}
\newblock
\APACjournalVolNumPages{Cosmic Rays and Particle Physics}{}{}{295}.
\PrintBackRefs{\CurrentBib}

\bibitem [\protect \citeauthoryear {%
G{\'o}mez%
}{%
G{\'o}mez%
}{%
{\protect \APACyear {2019}}%
}]{%
gomez2019muon}
\APACinsertmetastar {%
gomez2019muon}%
\begin{APACrefauthors}%
G{\'o}mez, H.%
\end{APACrefauthors}%
\unskip\
\newblock
\APACrefYearMonthDay{2019}{}{}.
\newblock
{\BBOQ}\APACrefatitle {Muon tomography using micromegas detectors: From
  Archaeology to nuclear safety applications} {Muon tomography using micromegas
  detectors: From archaeology to nuclear safety applications}.{\BBCQ}
\newblock
\APACjournalVolNumPages{Nuclear Instruments and Methods in Physics Research
  Section A: Accelerators, Spectrometers, Detectors and Associated
  Equipment}{936}{}{14--17}.
\PrintBackRefs{\CurrentBib}

\bibitem [\protect \citeauthoryear {%
Grieder%
}{%
Grieder%
}{%
{\protect \APACyear {2001}}%
}]{%
grieder2001cosmic}
\APACinsertmetastar {%
grieder2001cosmic}%
\begin{APACrefauthors}%
Grieder, P\BPBI K.%
\end{APACrefauthors}%
\unskip\
\newblock
\APACrefYear{2001}.
\newblock
\APACrefbtitle {Cosmic rays at {Earth}} {Cosmic rays at {Earth}}.
\newblock
\APACaddressPublisher{}{Elsevier}.
\PrintBackRefs{\CurrentBib}

\bibitem [\protect \citeauthoryear {%
Haino%
\ \protect \BOthers {.}}{%
Haino%
\ \protect \BOthers {.}}{%
{\protect \APACyear {2004}}%
}]{%
haino2004surfmuons}
\APACinsertmetastar {%
haino2004surfmuons}%
\begin{APACrefauthors}%
Haino, S.%
, Sanuki, T.%
, Abe, K.%
, Anraku, K.%
, Asaoka, Y.%
, Fuke, H.%
\BDBL {}others%
\end{APACrefauthors}%
\unskip\
\newblock
\APACrefYearMonthDay{2004}{}{}.
\newblock
{\BBOQ}\APACrefatitle {Measurements of primary and atmospheric cosmic-ray
  spectra with the {BESS-TeV} spectrometer} {Measurements of primary and
  atmospheric cosmic-ray spectra with the {BESS-TeV} spectrometer}.{\BBCQ}
\newblock
\APACjournalVolNumPages{Physics Letters B}{594}{1-2}{35--46}.
\PrintBackRefs{\CurrentBib}

\bibitem [\protect \citeauthoryear {%
Hersbach%
\ \protect \BOthers {.}}{%
Hersbach%
\ \protect \BOthers {.}}{%
{\protect \APACyear {2018}}%
}]{%
ERA5}
\APACinsertmetastar {%
ERA5}%
\begin{APACrefauthors}%
Hersbach, H.%
, Bell, B.%
, Berrisford, P.%
, Biavati, G.%
, Horányi, A.%
, Muñoz~Sabater, J.%
\BDBL {}Thépaut, J\BHBI N.%
\end{APACrefauthors}%
\unskip\
\newblock
\APACrefYearMonthDay{2018}{}{}.
\newblock
\APACrefbtitle {{ERA5} hourly data on pressure levels from 1979 to preset.}
  {{ERA5} hourly data on pressure levels from 1979 to preset.}
\newblock
\APACaddressPublisher{}{Copernicus Climate Change Service (C3S) Climate Data
  Store (CDS)}.
\newblock
\APACrefnote{Accessed on 30-11-2020}
\newblock
\begin{APACrefDOI} \doi{10.24381/cds.bd0915c6} \end{APACrefDOI}
\PrintBackRefs{\CurrentBib}

\bibitem [\protect \citeauthoryear {%
Kohno%
, Imai%
, Inue%
, Kodama%
\BCBL {}\ \BBA {} Wada%
}{%
Kohno%
\ \protect \BOthers {.}}{%
{\protect \APACyear {1981}}%
}]{%
kohno1981estimation}
\APACinsertmetastar {%
kohno1981estimation}%
\begin{APACrefauthors}%
Kohno, T.%
, Imai, K.%
, Inue, A.%
, Kodama, M.%
\BCBL {}\ \BBA {} Wada, M.%
\end{APACrefauthors}%
\unskip\
\newblock
\APACrefYearMonthDay{1981}{}{}.
\newblock
{\BBOQ}\APACrefatitle {Estimation of the vertical profile of atmospheric
  temperature from cosmic-ray components} {Estimation of the vertical profile
  of atmospheric temperature from cosmic-ray components}.{\BBCQ}
\newblock
\BIn{} \APACrefbtitle {International Cosmic Ray Conference} {International
  cosmic ray conference}\ (\BVOL~10, \BPG~289).
\PrintBackRefs{\CurrentBib}

\bibitem [\protect \citeauthoryear {%
Lipari%
\ \BBA {} Stanev%
}{%
Lipari%
\ \BBA {} Stanev%
}{%
{\protect \APACyear {1991}}%
}]{%
lipari1991}
\APACinsertmetastar {%
lipari1991}%
\begin{APACrefauthors}%
Lipari, P.%
\BCBT {}\ \BBA {} Stanev, T.%
\end{APACrefauthors}%
\unskip\
\newblock
\APACrefYearMonthDay{1991}{}{}.
\newblock
{\BBOQ}\APACrefatitle {Propagation of multi-{TeV} muons} {Propagation of
  multi-{TeV} muons}.{\BBCQ}
\newblock
\APACjournalVolNumPages{Physical Review D}{44}{11}{3543}.
\PrintBackRefs{\CurrentBib}

\bibitem [\protect \citeauthoryear {%
Lázaro~Roche%
}{%
Lázaro~Roche%
}{%
{\protect \APACyear {2021}}%
}]{%
particles4030028}
\APACinsertmetastar {%
particles4030028}%
\begin{APACrefauthors}%
Lázaro~Roche, I.%
\end{APACrefauthors}%
\unskip\
\newblock
\APACrefYearMonthDay{2021}{}{}.
\newblock
{\BBOQ}\APACrefatitle {A Compact Muon Tracker for Dynamic Tomography of Density
  Based on a Thin Time Projection Chamber with Micromegas Readout} {A compact
  muon tracker for dynamic tomography of density based on a thin time
  projection chamber with micromegas readout}.{\BBCQ}
\newblock
\APACjournalVolNumPages{Particles}{4}{3}{333--342}.
\newblock
\begin{APACrefDOI} \doi{10.3390/particles4030028} \end{APACrefDOI}
\PrintBackRefs{\CurrentBib}

\bibitem [\protect \citeauthoryear {%
Miyazaki%
\ \BBA {} Wada%
}{%
Miyazaki%
\ \BBA {} Wada%
}{%
{\protect \APACyear {1970}}%
}]{%
miyazaki}
\APACinsertmetastar {%
miyazaki}%
\begin{APACrefauthors}%
Miyazaki, Y.%
\BCBT {}\ \BBA {} Wada, M.%
\end{APACrefauthors}%
\unskip\
\newblock
\APACrefYearMonthDay{1970}{}{}.
\newblock
{\BBOQ}\APACrefatitle {Simulation of cosmic ray variation due to temperature
  effect} {Simulation of cosmic ray variation due to temperature
  effect}.{\BBCQ}
\newblock
\BIn{} \APACrefbtitle {International Cosmic Ray Conference} {International
  cosmic ray conference}\ (\BVOL~2, \BPG~591).
\PrintBackRefs{\CurrentBib}

\bibitem [\protect \citeauthoryear {%
Morishima%
\ \protect \BOthers {.}}{%
Morishima%
\ \protect \BOthers {.}}{%
{\protect \APACyear {2017}}%
}]{%
pyramids2017morishima}
\APACinsertmetastar {%
pyramids2017morishima}%
\begin{APACrefauthors}%
Morishima, K.%
, Kuno, M.%
, Nishio, A.%
, Kitagawa, N.%
, Manabe, Y.%
, Moto, M.%
\BDBL {}others%
\end{APACrefauthors}%
\unskip\
\newblock
\APACrefYearMonthDay{2017}{}{}.
\newblock
{\BBOQ}\APACrefatitle {Discovery of a big void in {Khufu’s} {Pyramid} by
  observation of cosmic-ray muons} {Discovery of a big void in {Khufu’s}
  {Pyramid} by observation of cosmic-ray muons}.{\BBCQ}
\newblock
\APACjournalVolNumPages{Nature}{552}{7685}{386--390}.
\PrintBackRefs{\CurrentBib}

\bibitem [\protect \citeauthoryear {%
Osprey%
\ \protect \BOthers {.}}{%
Osprey%
\ \protect \BOthers {.}}{%
{\protect \APACyear {2009}}%
}]{%
osprey2009suddenminos}
\APACinsertmetastar {%
osprey2009suddenminos}%
\begin{APACrefauthors}%
Osprey, S.%
, Barnett, J.%
, Smith, J.%
, Adamson, P.%
, Andreopoulos, C.%
, Arms, K.%
\BDBL {}others%
\end{APACrefauthors}%
\unskip\
\newblock
\APACrefYearMonthDay{2009}{}{}.
\newblock
{\BBOQ}\APACrefatitle {Sudden stratospheric warmings seen in {MINOS} deep
  underground muon data} {Sudden stratospheric warmings seen in {MINOS} deep
  underground muon data}.{\BBCQ}
\newblock
\APACjournalVolNumPages{Geophysical Research Letters}{36}{5}{}.
\PrintBackRefs{\CurrentBib}

\bibitem [\protect \citeauthoryear {%
Pla-Dalmau%
, Bross%
\BCBL {}\ \BBA {} Rykalin%
}{%
Pla-Dalmau%
\ \protect \BOthers {.}}{%
{\protect \APACyear {2003}}%
}]{%
pla2003extruding}
\APACinsertmetastar {%
pla2003extruding}%
\begin{APACrefauthors}%
Pla-Dalmau, A.%
, Bross, A\BPBI D.%
\BCBL {}\ \BBA {} Rykalin, V\BPBI V.%
\end{APACrefauthors}%
\unskip\
\newblock
\APACrefYearMonthDay{2003}{}{}.
\newblock
{\BBOQ}\APACrefatitle {Extruding plastic scintillator at {Fermilab}} {Extruding
  plastic scintillator at {Fermilab}}.{\BBCQ}
\newblock
\BIn{} \APACrefbtitle {2003 IEEE Nuclear Science Symposium. Conference Record
  (IEEE Cat. No. 03CH37515)} {2003 ieee nuclear science symposium. conference
  record (ieee cat. no. 03ch37515)}\ (\BVOL~1, \BPGS\ 102--104).
\PrintBackRefs{\CurrentBib}

\bibitem [\protect \citeauthoryear {%
Procureur%
}{%
Procureur%
}{%
{\protect \APACyear {2018}}%
}]{%
procureur2018muonimaging}
\APACinsertmetastar {%
procureur2018muonimaging}%
\begin{APACrefauthors}%
Procureur, S.%
\end{APACrefauthors}%
\unskip\
\newblock
\APACrefYearMonthDay{2018}{}{}.
\newblock
{\BBOQ}\APACrefatitle {Muon imaging: Principles, technologies and applications}
  {Muon imaging: Principles, technologies and applications}.{\BBCQ}
\newblock
\APACjournalVolNumPages{Nuclear Instruments and Methods in Physics Research
  Section A: Accelerators, Spectrometers, Detectors and Associated
  Equipment}{878}{}{169--179}.
\PrintBackRefs{\CurrentBib}

\bibitem [\protect \citeauthoryear {%
Reyna%
}{%
Reyna%
}{%
{\protect \APACyear {2006}}%
}]{%
reyna2006simple}
\APACinsertmetastar {%
reyna2006simple}%
\begin{APACrefauthors}%
Reyna, D.%
\end{APACrefauthors}%
\unskip\
\newblock
\APACrefYearMonthDay{2006}{}{}.
\newblock
{\BBOQ}\APACrefatitle {A simple parameterization of the cosmic-ray muon
  momentum spectra at the surface as a function of zenith angle} {A simple
  parameterization of the cosmic-ray muon momentum spectra at the surface as a
  function of zenith angle}.{\BBCQ}
\newblock
\APACjournalVolNumPages{arXiv preprint hep-ph/0604145}{}{}{}.
\PrintBackRefs{\CurrentBib}

\bibitem [\protect \citeauthoryear {%
Ri{\'a}digos%
, Garc{\'\i}a-Castro%
, Gonz{\'a}lez-D{\'\i}az%
\BCBL {}\ \BBA {} P{\'e}rez-Mu{\~n}uzuri%
}{%
Ri{\'a}digos%
\ \protect \BOthers {.}}{%
{\protect \APACyear {2020}}%
}]{%
riadigos2020atmospheric}
\APACinsertmetastar {%
riadigos2020atmospheric}%
\begin{APACrefauthors}%
Ri{\'a}digos, I.%
, Garc{\'\i}a-Castro, D.%
, Gonz{\'a}lez-D{\'\i}az, D.%
\BCBL {}\ \BBA {} P{\'e}rez-Mu{\~n}uzuri, V.%
\end{APACrefauthors}%
\unskip\
\newblock
\APACrefYearMonthDay{2020}{}{}.
\newblock
{\BBOQ}\APACrefatitle {Atmospheric Temperature Effect in Secondary Cosmic Rays
  Observed With a 2 m2 Ground-Based t{RPC} Detector} {Atmospheric temperature
  effect in secondary cosmic rays observed with a 2 m2 ground-based t{RPC}
  detector}.{\BBCQ}
\newblock
\APACjournalVolNumPages{Earth and Space Science}{7}{9}{e2020EA001131}.
\PrintBackRefs{\CurrentBib}

\bibitem [\protect \citeauthoryear {%
Rockenbach%
\ \protect \BOthers {.}}{%
Rockenbach%
\ \protect \BOthers {.}}{%
{\protect \APACyear {2014}}%
}]{%
GMDNrockenbach}
\APACinsertmetastar {%
GMDNrockenbach}%
\begin{APACrefauthors}%
Rockenbach, M.%
, Dal~Lago, A.%
, Schuch, N.%
, Munakata, K.%
, Kuwabara, T.%
, Oliveira, A.%
\BDBL {}others%
\end{APACrefauthors}%
\unskip\
\newblock
\APACrefYearMonthDay{2014}{}{}.
\newblock
{\BBOQ}\APACrefatitle {Global muon detector network used for space weather
  applications} {Global muon detector network used for space weather
  applications}.{\BBCQ}
\newblock
\APACjournalVolNumPages{Space Science Reviews}{182}{1-4}{1--18}.
\PrintBackRefs{\CurrentBib}

\bibitem [\protect \citeauthoryear {%
Rodgers%
}{%
Rodgers%
}{%
{\protect \APACyear {2000}}%
}]{%
rodgers2000inverse}
\APACinsertmetastar {%
rodgers2000inverse}%
\begin{APACrefauthors}%
Rodgers, C\BPBI D.%
\end{APACrefauthors}%
\unskip\
\newblock
\APACrefYear{2000}.
\newblock
\APACrefbtitle {Inverse methods for atmospheric sounding: theory and practice}
  {Inverse methods for atmospheric sounding: theory and practice}\ (\BVOL~2).
\newblock
\APACaddressPublisher{}{World scientific}.
\PrintBackRefs{\CurrentBib}

\bibitem [\protect \citeauthoryear {%
Sagisaka%
}{%
Sagisaka%
}{%
{\protect \APACyear {1986}}%
}]{%
sagisaka1986}
\APACinsertmetastar {%
sagisaka1986}%
\begin{APACrefauthors}%
Sagisaka, S.%
\end{APACrefauthors}%
\unskip\
\newblock
\APACrefYearMonthDay{1986}{}{}.
\newblock
{\BBOQ}\APACrefatitle {Atmospheric effects on cosmic-ray muon intensities at
  deep underground depths} {Atmospheric effects on cosmic-ray muon intensities
  at deep underground depths}.{\BBCQ}
\newblock
\APACjournalVolNumPages{Il Nuovo Cimento C}{9}{4}{809--828}.
\PrintBackRefs{\CurrentBib}

\bibitem [\protect \citeauthoryear {%
Tilav%
\ \protect \BOthers {.}}{%
Tilav%
\ \protect \BOthers {.}}{%
{\protect \APACyear {2010}}%
}]{%
tilav2010}
\APACinsertmetastar {%
tilav2010}%
\begin{APACrefauthors}%
Tilav, S.%
, Desiati, P.%
, Kuwabara, T.%
, Rocco, D.%
, Rothmaier, F.%
, Simmons, M.%
\BCBL {}\ \BBA {} Wissing, H.%
\end{APACrefauthors}%
\unskip\
\newblock
\APACrefYearMonthDay{2010}{}{}.
\newblock
{\BBOQ}\APACrefatitle {Atmospheric Variations as observed by {IceCube}}
  {Atmospheric variations as observed by {IceCube}}.{\BBCQ}
\newblock
\APACjournalVolNumPages{arXiv preprint arXiv:1001.0776}{}{}{}.
\PrintBackRefs{\CurrentBib}

\bibitem [\protect \citeauthoryear {%
Trzaska%
\ \protect \BOthers {.}}{%
Trzaska%
\ \protect \BOthers {.}}{%
{\protect \APACyear {2019}}%
}]{%
trzaska2019canfranc}
\APACinsertmetastar {%
trzaska2019canfranc}%
\begin{APACrefauthors}%
Trzaska, W\BPBI H.%
, Slupecki, M.%
, Bandac, I.%
, Bayo, A.%
, Bettini, A.%
, Bezrukov, L.%
\BDBL {}others%
\end{APACrefauthors}%
\unskip\
\newblock
\APACrefYearMonthDay{2019}{}{}.
\newblock
{\BBOQ}\APACrefatitle {Cosmic-ray muon flux at Canfranc Underground Laboratory}
  {Cosmic-ray muon flux at canfranc underground laboratory}.{\BBCQ}
\newblock
\APACjournalVolNumPages{The European Physical Journal C}{79}{8}{1--5}.
\PrintBackRefs{\CurrentBib}

\bibitem [\protect \citeauthoryear {%
Watanabe%
\ \protect \BOthers {.}}{%
Watanabe%
\ \protect \BOthers {.}}{%
{\protect \APACyear {2019}}%
}]{%
watanabe2019compensated}
\APACinsertmetastar {%
watanabe2019compensated}%
\begin{APACrefauthors}%
Watanabe, K.%
, Tanaka, S.%
, Chang, W.%
, Chen, H.%
, Chu, M.%
, Cuenca-Garc{\'\i}a, J.%
\BDBL {}others%
\end{APACrefauthors}%
\unskip\
\newblock
\APACrefYearMonthDay{2019}{}{}.
\newblock
{\BBOQ}\APACrefatitle {A compensated multi-gap {RPC} with 2 m strips for the
  {LEPS2} experiment} {A compensated multi-gap {RPC} with 2 m strips for the
  {LEPS2} experiment}.{\BBCQ}
\newblock
\APACjournalVolNumPages{Nuclear Instruments and Methods in Physics Research
  Section A: Accelerators, Spectrometers, Detectors and Associated
  Equipment}{925}{}{188--192}.
\PrintBackRefs{\CurrentBib}

\bibitem [\protect \citeauthoryear {%
Xing-Ming%
\ \protect \BOthers {.}}{%
Xing-Ming%
\ \protect \BOthers {.}}{%
{\protect \APACyear {2014}}%
}]{%
xing2014position}
\APACinsertmetastar {%
xing2014position}%
\begin{APACrefauthors}%
Xing-Ming, F.%
, Yi, W.%
, Xue-Wu, W.%
, Zhi, Z.%
, Ming, Z.%
, Zi-Ran, Z.%
\BDBL {}Gonzalez-Diaz, D.%
\end{APACrefauthors}%
\unskip\
\newblock
\APACrefYearMonthDay{2014}{}{}.
\newblock
{\BBOQ}\APACrefatitle {A position resolution {MRPC} for muon tomography} {A
  position resolution {MRPC} for muon tomography}.{\BBCQ}
\newblock
\APACjournalVolNumPages{Chinese Physics C}{38}{4}{046003}.
\PrintBackRefs{\CurrentBib}

\bibitem [\protect \citeauthoryear {%
Yanchukovsky%
}{%
Yanchukovsky%
}{%
{\protect \APACyear {2020}}%
}]{%
yanchukovsky2020}
\APACinsertmetastar {%
yanchukovsky2020}%
\begin{APACrefauthors}%
Yanchukovsky, V.%
\end{APACrefauthors}%
\unskip\
\newblock
\APACrefYearMonthDay{2020}{}{}.
\newblock
{\BBOQ}\APACrefatitle {Muon intensity variations and atmospheric temperature}
  {Muon intensity variations and atmospheric temperature}.{\BBCQ}
\newblock
\APACjournalVolNumPages{Solar-Terrestrial Physics}{6}{1}{108--115}.
\PrintBackRefs{\CurrentBib}

\bibitem [\protect \citeauthoryear {%
Yanchukovsky%
, Filimonov%
\BCBL {}\ \BBA {} Hisamov%
}{%
Yanchukovsky%
\ \protect \BOthers {.}}{%
{\protect \APACyear {2007}}%
}]{%
yanchukovsky2007}
\APACinsertmetastar {%
yanchukovsky2007}%
\begin{APACrefauthors}%
Yanchukovsky, V.%
, Filimonov, G\BPBI Y.%
\BCBL {}\ \BBA {} Hisamov, R.%
\end{APACrefauthors}%
\unskip\
\newblock
\APACrefYearMonthDay{2007}{}{}.
\newblock
{\BBOQ}\APACrefatitle {Atmospheric variations in muon intensity for different
  zenith angles} {Atmospheric variations in muon intensity for different zenith
  angles}.{\BBCQ}
\newblock
\APACjournalVolNumPages{Bulletin of the Russian Academy of Sciences:
  Physics}{71}{7}{1038--1040}.
\PrintBackRefs{\CurrentBib}

\bibitem [\protect \citeauthoryear {%
Yanchukovsky%
, Sunyakov%
\BCBL {}\ \BBA {} Kuzmenko%
}{%
Yanchukovsky%
\ \protect \BOthers {.}}{%
{\protect \APACyear {2015}}%
}]{%
yanchukovsky2015}
\APACinsertmetastar {%
yanchukovsky2015}%
\begin{APACrefauthors}%
Yanchukovsky, V.%
, Sunyakov, S.%
\BCBL {}\ \BBA {} Kuzmenko, V.%
\end{APACrefauthors}%
\unskip\
\newblock
\APACrefYearMonthDay{2015}{}{}.
\newblock
{\BBOQ}\APACrefatitle {Variations in temperature at different isobaric levels
  of the atmosphere, according to data on cosmic rays} {Variations in
  temperature at different isobaric levels of the atmosphere, according to data
  on cosmic rays}.{\BBCQ}
\newblock
\APACjournalVolNumPages{Bulletin of the Russian Academy of Sciences:
  Physics}{79}{5}{667--669}.
\PrintBackRefs{\CurrentBib}

\end{thebibliography}

%
%
%
%
%

\end{document}